\documentclass{aastex}

\def\eg{{\it e.g.\ }}
\def\etal{{\it et al.\thinspace}}

\begin{document}

\title{The Relation Between Activity and Environment in Compact Groups of Galaxies}

\author{Roger Coziol\altaffilmark{1}, Angela Iovino\altaffilmark{1}}

\and

\author{Reinaldo R. de Carvalho\altaffilmark{2}}

\altaffiltext{1}{Osservatorio Astronomico di Brera, Via Brera 28,
I-20121 Milano, Italy} \altaffiltext{2}{Observat\'orio Nacional,
Rua Gal.  Jos\'e Cristino, 77 -- 20921-400, Rio de Janeiro, RJ.,
Brasil}

\begin{abstract}

We present the results of the classification of spectral activity
types for 193 galaxies from a new sample of 49 compact groups of
galaxies in the southern hemisphere (SCGs).  The SCGs is a new
sample of compact groups selected in automated fashion from a
digitized galaxy catalogue, covering an area of $\sim 5200
~sq~deg$ around the South Galactic Pole. It is complete up to
m$\sim 14.5$ in $b_j$ for the brightest galaxy of the group. This
sample is very well suited for statistical studies of compact
groups properties as it is unaffected by the biases introduced by
visual selection methods.

The spectral analysis of the SCG galaxies confirms the results
previously obtained from the observation of a smaller sample of 17
CGs in Hickson's sample (HCG). We confirm the luminosity--activity
and morphology--activity relations previously observed in HCGs: 1)
low luminosity AGNs and AGNs are preferentially located in the
most luminous galaxies in the groups, while non--emission--line
galaxies and star--forming galaxies share a common distribution
among the less luminous galaxies; 2) the non--emission--line
galaxies, the low luminosity AGNs and AGNs are nearly all in
early--type galaxies, while the star--forming galaxies are
preferentially located in late--type spirals.  We also verified
that the number of ``evolved'' galaxies (early--type
non--emission--line galaxies) significantly increases with the
number of members in the group. Finally, we confirm that AGNs
(including low luminosity or dwarf AGNs) are the most frequent
(41\%) activity type encountered in CGs.

The SCGs contain more star--forming galaxies and less
non--emission--line galaxies than HCGs. This difference is
possibly related to the different selection criteria of SCGs with
respect to HCGs. Being selected in an automated way, the SCGs
probe a wider range of physical properties than the HCGs, which
are biased towards more easily detectable, denser and therefore
more evolved groups.

The star--forming galaxies in SCGs are composed of few starburst
galaxies and a higher number of \ion{H}{2} Nucleus Galaxies, which
generally have less intense star formation than starburst
galaxies. As a consequence, the star formation activity in SCGs is
remarkably low. At the same time, we found clear evidence for
nuclear activity in all the star--forming galaxies.  Adding up the
percentage of AGNs and star--forming galaxies showing evidence of
nuclear activity would result in more than 70\% of the galaxies in
SCGs having an active nucleus.  If these results are further
confirmed, CGs will be in the local Universe the best location
where to find AGNs. Curiously, however, this characteristic of CGs
generally excludes Seyfert 1 galaxies.

\end{abstract}


\section{Introduction}

Studies on Compact Groups (CGs) of galaxies mainly concentrate on
two issues: 1) what is the origin and relative importance of CGs
in the Universe? 2) is there a relation between the makeup of
these structures and the formation and evolution of galaxies which
form them?  Actually, the possibility that the two issues are
connected is what makes CGs so interesting in the first place (see
the review by Hickson 1997).  Now that we have some unambiguous
evidence of galaxy evolution driven by dynamical environment
effects, even at relatively modest redshifts (\eg O'Connell 1999,
VanDokkum et al. 1999), CGs with their high spatial densities and
small velocity dispersions look more than ever like the ideal
objects to investigate how dynamical effects forge galaxy
evolution.

Although the first dynamical simulations of CGs predicted that
they should evolve within only a few orbital times into a system
dominated by one central merger remnant (Barnes 1989), evidence
for on--going or post merger events in CGs are scarce
(Mardirossian et al.\ 1982; Zepf \& Whitmore 1991; Zepf 1993;
Moles et al.\ 1994).  The high number of these systems also
suggests longer lifetime for CGs, as compared to what is expected
based on the small crossing times of the galaxies. But, these
contradictions may only be apparent, due to our still limited
knowledge of how mergers really proceed and what kind of remnants
they leave behind (see the review by Mihos 1999). This is the
point made by several recent numerical simulations, which show
that the evolution of CGs may take more than a few Gyrs if they
are embedded in massive dark matter halos (Athanassoula et al.
1997) or even longer if the merging galaxies had originally a wide
range in masses (Governato et al. 1991).

Our difficulties in understanding CGs may also be a consequence of
our lack of knowledge of what these structures really are. Short
dynamical lifetimes and a high number of CGs, for instance, are
not a problem if these structures are collapsing physical systems
continuously forming in loose groups (Diaferio et al.\ 1994;
Governato et al. 1996). This seems to be the converging conclusion
of many recent redshift studies of galaxies around compact groups
(Garcia 1995; Ribeiro et al.\ 1998; Barton et al.\ 1998). In
Ribeiro et al.\ (1998), in particular, it was shown that amongst
CGs there is a wide range of different dynamical stages, with
different surface density profiles; the core radius increasing
from values typical of very compact to loose groups. These authors
confirmed that the fraction of spiral galaxies is lower in CGs
than in the field and that it decreases as the velocity dispersion
increases (Hickson et al.\ 1988), suggesting that the environment
influenced the formation and/or evolution of the galaxies in the
groups. In these structures, interactions and mergers are
suggested by the decrease of spiral fraction with the decrease of
crossing time and increase of X-ray luminosity (Ponman et al.\
1996). The linear relation between velocity dispersion and gas
temperature also seems to get steeper in CGs than in cluster,
which is consistent with effects of dynamical friction (Bird et
al. 1995).

The dynamical evidences in favor of environmental effects, as
described above, are consistent with our previous analysis of
activity types found in HCGs (Ribeiro et al.\ 1996; Coziol et al.\
1998a, 1998b; de Carvalho \& Coziol 1999). We have finds that a
significant fraction of the brightest galaxies in the groups
displays some sort of activity, either star formation or AGN. We
also discovered the presence of a large number of low luminosity
AGNs, making AGN the dominant activity type in CGs (Coziol et al.\
1998a). Related to this activity, we find a
density--morphology--activity relation, according to which the
AGNs (low luminosity AGNs included) and galaxies without emission
lines are the most luminous and invariably early-type galaxies and
tend to concentrate in the core of HCGs, while the star--forming
galaxies, which may also be preferentially late--type spirals, are
more numerous in the halo (Coziol et al.\ 1998b).

Despite looking like normal field galaxies with similar
morphologies, galaxies in CGs seem to have stellar metallicity
that is relatively high for their luminosity. They also have
slightly narrower metal absorption line equivalent widths and
relatively strong Balmer absorption lines which suggests that a
meager extra population of intermediate--age stars is still
present (Coziol et al.  1998b).  This phenomenon may be
reminiscent of the ``post--starburst'' phase associated to the
Butcher--Oemler effect (Belloni et al.\ 1995; Caldwell \& Rose
1997). But, the ``post--starburst'' signs in CG galaxies look
weaker than usually observed (Poggianti \& Barbaro 1996; Zabludoff
et al. 1996; Leonardi \& Rose 1996), which suggests a more
advanced stage of evolution (or milder level of past star--forming
activity) for the galaxies in the groups (Caon et al.\ 1994;
Coziol et al.\ 1998b).

Considering the significance of our results for understanding how
the environment of galaxies affects their formation and evolution,
we believe that it is important to confirm our previous
observations on the activity in CGs on a larger and complete
sample.  This is the main subject of the present article.  In
Section~2, we present the new sample of CG galaxies used for our
analysis and give some details on their spectroscopic
observations.  In Section~3, we discuss our classification method
and present the results. A summary of our study and conclusions
can be found in Section~4.

\section{A new sample of CGs in the southern sky}

\subsection{Characteristics of the sample}

With the aim of obtaining a larger and deeper sample of CG of
galaxies, Iovino \& Tassi (1999; hereafter IT99), developed an
algorithm to search for compact groups of galaxies on digitized
galaxy catalog.  The selection criteria used by these authors are
slightly different than those used by Hickson: the richness and
compactness criteria are maintained, while the isolation criterion
is relaxed.  In formulae, the conditions are: 1 -- richness: $n
\ge 4$ with $m\leq m_F$ and $ m_B \leq m_F \leq m_B+3 $; 2 --
compactness: $\mu_G < 27.7$; 3 -- isolation: $R_N \ge 3R_G$. Where
$n$ is the number of member galaxies, $m_B$ the estimated blue
$b_j$ magnitude of the brightest group member, $m_F$ the estimated
blue $b_j$ magnitude of the faintest group member (at most equal
to $m_B + 3$), $R_G$ the group radius, corresponding to the
smallest circle containing the centers of the group members, and
$R_N$ the distance from the center of the group to the nearest non
member galaxy within the same magnitude range or brighter. $\mu_G$
is the surface brightness of the group (mag per arcsec$^2$)
defined as the total magnitude of all the member galaxies averaged
over the circle with radius $R_N$.  Changing from $b_j$ to E, this
threshold is equivalent to the one used by Hickson, namely 26.0
magnitude per arcsec$^2$.  As a further constraint, it is
requested that the mean surface brightness in the isolation
annulus (now including galaxies up to 3 magnitudes fainter than
the brightest group member) should be fainter than $\mu_G + 4$.
Having relaxed the isolation constraint, this last measure is
necessary to avoid the selection of many sub-condensations within
larger structures, \eg clusters.

Applying this algorithm to the SERC(J) Southern Sky Survey,
digitized by the COSMOS plate scanning machine, IT99 produced a
list of 60 Southern CGs (SCGs) with magnitude of the starting
galaxy brighter than 14.5 in $b_j$ on an area of $\sim 5200
~sq~deg$ around the South Galactic Pole. The spectroscopic
follow--up of this sample resulted in the confirmation of 49
groups of galaxies with three or more concordant members. A
velocity cut--off of 1000 km/sec from the median velocity of the
group was assumed.

In this paper we present the results of the classification of
spectral activity types for the 193 galaxies of the 49 concordant
groups. Being selected in an automated fashion, such sample is
extremely well suited for statistical studies of compact groups
properties, as it is unaffected by the biases introduced by visual
selection methods (Prandoni \etal 1994).

\subsection {Spectroscopic observations}

The material we used for our study is the same material presented
in Iovino \etal 1999 (I99), \eg the one collected during the
spectroscopic follow--up of SCGs.  One should refer to I99 for a
detailed log of the observations. Here, it will be enough to
remind that the spectroscopic data were all obtained at the 1.5m
ESO telescope of La Silla, Chile, during a series of three
observing runs in 1995 (four nights), 1996 (three nights) and 1997
(five nights).  The Boller\&Chivens spectrograph was the
instrument used, with different CCDs and gratings.  In 1995 we
used the $2048 \times 2048$ pixels, CCD\#24. During the 1996 and
1997 runs we used instead CCD \#39, a Loral/Lesser $2048 \times
2048$ CCD, whose is twice as efficient as CCD \#24 in the red part
of the spectrum, and five times as efficient in the blue end. Both
CCD's have pixel size of $15 \mu$. During observations, they were
windowed in the spatial direction for faster readout.

Except for the first night of 1995, the grating used was always
ESO \# 23. This grating, blazed at $4550$\AA, yields a pixel size
of $0\farcs82 \times 1.9$\AA\ and gives a wavelength coverage of
$3700-7600$\AA.  The first night of 1995 run, the grating ESO \#2
with a disperson of $ 300 \,$l/mm was used. Blazed at $4550$\AA,
it gives a pixel size of $ 0\farcs82 \times 3.8$\AA\ and a
wavelength coverage of $ 3600-10500$\AA. The slit width was always
set to 2$\farcs$ throughout the runs, resulting in a spectral
resolution of $\sim 4.6 $\AA ~for the spectra taken with grating
ESO \#23, and $\sim 9.3 $\AA ~for the spectra taken with grating
ESO \#2. Typical exposure times varied from 10 to 30 minutes
depending on the magnitude of the galaxy to be observed. Because
they were taken for kinematics studies, the spectra were not
calibrated in flux.

Data reduction employed standard routines available in {\tt
IRAF}~\footnote{{\tt IRAF} is distributed by the National Optical
Astronomy Observatories, which are operated by the Association of
Universities for Research in Astronomy, Inc. under contract with
the National Science Foundation}.  Following subtraction of bias
and dark, flat field correction, and cosmic rays removal, one
dimensional spectra were extracted from each frame using the
variance weighting option in the task {\tt APALL}.

\section{Classification of activity types}

\subsection{General classification: spectral characteristics
of the activity types}

The classification of activity types of the SCG galaxies was done
in two steps.  First we identified galaxies with and without line
emission. Then, for the emission--line galaxies, we measured
various emission line ratios in order to build different standard
diagnostic diagrams (Baldwin et al.  1981; Veilleux \& Osterbrock
1987).  As Veilleux et al. (1995) have shown, it is useful to use
more than one diagnostic diagrams to identify ``intermediate'' or
ambiguous cases, \eg galaxies where a weak AGN or gas excited by
shocks could be mixed with star formation (for more details on
this subject see Coziol et al.  1999).

We illustrate our spectral classification by showing in Figures~1
to 3 examples of galaxies with different activity types.  We
separate the spectra into three categories: galaxies without
emission, emission--line galaxies (AGNs, LINERs and star forming
galaxies) and low--luminosity AGNs.  Some spectra of
non--emission--line galaxies are presented in Figure~1a.  The
spectra of these galaxies are similar to those of elliptical and
early--type spiral galaxies.  They show a mixture of old and
intermediate age stellar populations, as judged by the strength of
the Balmer lines in absorption and the amplitude of the Balmer
jump around 4000\AA.

Among the emission--line galaxies, we distinguish between galaxies
with star formation in their nucleus and galaxies where the gas is
excited by an AGN (LINER and Seyfert 2).  The spectral
characteristic which separates AGNs from star forming galaxies is
more intense forbidden lines, like [\ion{N}{2}]$\lambda6584$,
[\ion{S}{2}]$\lambda\lambda6717,6731$ and
\ion{O}{1}]$\lambda6300$. LINERs (Figure~2 a-b) are discriminated
from Seyfert 2 galaxies (Figure 2e-f) by their low level of
excitation.  In many cases, however, the differentiation between
these two types of AGNs is not obvious (Figure 2c-d) and can be
made only after comparing different line ratios using standard
diagnostic diagrams (see next section).

We present in Figure~3, some spectra of star forming galaxies in
our sample.  These galaxies usually show a featureless continuum
and strong emission lines. In the SCGs, we distinguish two types
of star forming galaxies.  The first type (Figure~3 a-b) have a
high excitation level ([\ion{O}{3}]$\lambda5007 $ H$\beta$) and a
relatively low ratio [\ion{N}{2}]$\lambda6584$/H$\alpha$, similar
to what is observed in normal \ion{H}{2} regions in the disk of
late--type spirals or in irregular galaxies.  The strong intensity
of the H$\alpha$ line suggests that they experience a relatively
high level of star formation.  The second type of star forming
galaxies have a low level of excitation and weak H$\alpha$
emission, which suggests that they have lower star formation than
the other star--forming galaxies (Figure~3 c-d).  The second type
of star forming galaxies also show unusually intense
[\ion{N}{2}]$\lambda6584$ and
[\ion{S}{2}]$\lambda\lambda6717,6731$ as compared to H$\alpha$.
As we will show later, these spectral characteristics are specific
to \ion{H}{2} Nucleus Galaxies (Kennicutt, Keel \& Blaha 1989;
Coziol 1996; Ho et al.  1998). The discrimination between star
forming galaxies and LINERs is not always obvious, and can be
achieved only using standard diagnostic diagrams. This is
particularly true for \ion{H}{2} Nucleus Galaxies, as we will see
in the next section.

A significant fraction of galaxies in SCGs cannot be easily
classified.  The spectra of some of these galaxies are presented
in Figure~1 c-d.  They show spectral characteristics which are
similar to those of non--emission--line galaxies (compare Figure 1
a with Figure~1 c), with the difference that they also show the
presence of some emission lines.  The more intense of these lines,
and in many cases the only one visible, is
[\ion{N}{2}]$\lambda6584$.  This phenomenon was already observed
in CGs and was shown to be related to the low--luminosity AGN
phenomenon (Coziol et al. 1998a). In a galaxy with a
low--luminosity AGN, the faint emission components originating
from the nucleus are severely diluted by absorption features
produced by intermediate mass stars present in the bulge. The
resulting emission line spectra of these galaxies is one where the
two principal Balmer emission lines are notably weak: H$\beta$
usually appears in absorption, while H$\alpha$ sometimes appears
in emission, but always with an apparent intensity ratio
[\ion{N}{2}]$\lambda6584$/H$\alpha > 1$.  In Coziol et al.
(1998a) we have demonstrated that after subtracting a template to
eliminate the dilution effect, these galaxies are invariably
classified as AGNs (Seyfert or LINERs, but mostly LINERs) with a
low luminosity (Filippenko \& Sargent 1985; Phillips et al.
1986).

Examining the optical spectral atlas of Ho et al. (1995), one can
verify that this phenomenon is very common in the nuclei of
luminous nearby galaxies.  The spectral characteristics of this
class of objects are well defined (see description above) and easy
to recognized by a trained eye. As a test, we verified that all
the galaxies in the spectral atlas presented by Ho et al. (1995)
selected as low--luminosity AGNs candidates by just applying the
phenomenological description given above were actually classified
as such by Ho et al. (1998) after template subtraction. Based on
this verification and on our previous experience dealing with
low--luminosity AGNs in CGs (Coziol et al. 1998a), we therefore
feel confident in classifying also as low--luminosity AGNs the
small fraction of galaxies (22\%) in our sample showing the same
typical spectral pattern (the low signal to noise ratios of our
spectra do not allow us to subtract a template).

The SCG galaxies with their activity types are presented in
Table~1. The whole sample is constituted of 193 galaxies
distributed in 49 CGs with 3 to 6 members. Following the name of
the groups, we give for each galaxy its blue ($b_j$) COSMOS
magnitude and its morphology. Errors in magnitude in the COSMOS
catalogue are smaller than 0.2 down to $b_j \sim 20$ (MacGillivray
\& Yentis 1993). The morphologies were determined by visual
inspection of our own R band CCD images of these groups. The
activity types are described as the following: non--emission--line
galaxies (no em.), low--luminosity AGNs (LLAGNs), LINERs, Seyfert
2 (Sy2) and star forming galaxies (SFGs). Our classification as
LINERs, Seyfert 2 or SFGs was established using three standard
diagnostic diagrams (next section).  A small number of galaxies in
our sample show emission lines similar to SFGs, but cannot be
classified because their spectra do not have a sufficiently high
S/N ratio (generally in the blue part of the spectra). In Table~1,
they are identified as EMLGs.

It is remarkable that we find mostly LINERs and Seyfert 2 galaxies
in our sample.  The only galaxies which resemble a Seyfert~1 is
SCG 0421-5624B.  The spectrum of this galaxy is shown in Figure~4.
The H$\alpha$ emission line may have a wider base ($\sim 80$ \AA,
see Figure~4c).  This large component, however, is not obvious in
H$\beta$ (cf.  Figure~2b).  We note also that the two forbidden
lines, [\ion{O}{3}]$\lambda\lambda4959,5007$ may have the same
width as H$\beta$, a characteristic proper to a Seyfert 2. This
observation is not conclusive, however, because of a peculiar
double peaks observed in these lines. This feature appear in three
different spectra taken in different nights, which suggests that
it is real. It is possible that we are seeing more than one
emission region in this galaxy. In this case, the superposition of
these regions would artificially enlarge the forbidden lines in
our spectrum and gives an erroneous identification as a Seyfert 2.
More observations are needed to settle this question.

\subsection{Classification from emission line diagnostic diagrams:
the ambiguous nature of SFGs in compact groups}

For classification purposes, we measured up to six different line
ratios.  Using the package SPLOT in IRAF, the intensities of the
lines were determined by fitting a gaussian. The level of the
continuum was estimated using two regions $\sim$ 100 \AA\ wide on
each side of the line. For the emission lines which are slightly
blended, like H$\alpha$ + [\ion{N}{2}] and the two sulphur lines,
the intensities were determined after fitting gaussians using the
``deblend'' option in SPLOT.

The measurements of emission line ratios are sensitive to the
dilution introduced by Balmer absorption lines, produced by
intermediate--age stellar populations present in the bulge of the
galaxies. To correct for this dilution effect, a template galaxy
spectrum is usually subtracted from the spectra (Veilleux et al.\
1995; Ho et al.\ 1997; Coziol et al.\ 1998a). However, the
relatively low signal to noise ratios of our spectra impede us to
apply this technique in the present study. To correct for this
effect, we used, therefore, an alternative method, which consists
in adding a constant equivalent width to the equivalent widths of
all the emission lines measured (see Contini et al. 1995 for an
explanation). To determine the EW value to add to our galaxies and
to verify the accuracy of this method, we first applied it to the
sample of HCG emission--line galaxies previously studied by Coziol
et al. (1998a) and compared the results with those previously
obtained after a template subtraction. We verified that by adding
a constant EW of 1 \AA, we can reproduce the results obtained
after template subtraction. In AGNs, however, we found that the
correction for EW(H$\beta$) implied by the standard correction
method is unrealistically high. This effect is caused by the
extreme narrowness of EW(H$\beta$) observed in these galaxies.
This happens because in some cases (mostly in Seyfert 2) H$\beta$
is so severely affected by dilution that we cannot measure it
correctly. The badly measured H$\beta$ line yields an unusually
narrow EW, which automatically translates into a large correction
factor (the correction factor varies as 1/EW). To overcome this
problem, we used an empirical relation to deduce the observed EW
in H$\beta$ from the one observed in H$\alpha$ (Sodr\'e and
Stasi\'nska 1999). We verified that this relation works well both
for AGNs and star forming galaxies. For consistence sake,
therefore, we decided to use it to deduce the observed value of
EW(H$\beta$) in all the emission line galaxies in our sample.

The corrected emission line ratios are presented in Table~2. The
uncertainties ($1\sigma$) on these values were determined assuming
Poisson statistics and applying standard error propagation
formula. From these measurements we deduce, in Table~3, the ratios
H$\alpha$/H$\beta$ from which we estimate the total
(Galactic$+$internal) extinction, E(B-V), using the Whitford
reddening curve as parametrized by Miller and Mathews (1972). We
adopted an intrinsic H$\alpha$/H$\beta$ ratio of 2.86 for SFGs and
3.10 for AGNs. The extinction corrected emission--line ratios are
given in columns 4 to 9 followed by the EW in H$\alpha$ in column
10.

A detailed classification of the activity types is given in the
last column of Table~3. The main types are: star forming galaxies
(H), LINER (L) or Seyfert (Sy).  To determine the activity type of
our galaxies we followed the same criteria as Ho et al. (1997a).
The logarithm of the line ratios adopted to distinguish between
the different activity types are listed in Table~4.  Generally we
give the classification suggested by more than one line ratios.
When the classification is ambiguous, we usually adopt the
activity type suggested by [\ion{O}{1}]$\lambda6300$/H$\alpha$ or
by [\ion{S}{2}]($\lambda6717+\lambda6731$)/H$\alpha$, when
[\ion{O}{1}] is not measured. In Table~3, a ``?'' indicates that
our classification is tentative, because it is based on only one
line ratio.  A ``:'' indicates an ambiguous nature for the galaxy,
that is, galaxy with different classifications in different
diagnostic diagrams. The high number of ambiguous cases in our
sample of emission--line galaxies is quite remarkable (31\% of our
galaxies, as compared to 19\% in the sample of Veilleux et al.
1995). In part, this ambiguity may be due to the method we used to
correct for the contamination by the stellar bulge population.
This method was shown to work well when applied to galaxies with
intense star formation like in disk HII regions and starburst
galaxies (Mc Call, Rybsky \& Shields 1985; Liu \& Kennicutt 1995;
Veilleux et al. 1995; Contini et al. 1998). However, its usage can
be more problematic when star formation is at a lower level than
in these galaxies or there maybe more than one source of
ionization of the gas, like in Seyfert and LINER galaxies. We
examine these questions further below.

The results of our classification of SCG galaxies in terms of the
three diagnostic diagrams are shown in Figure~5.  One can see
(Figure~5a) that some SFGs and LINERs have nitrogen ratios which
are different than usually expected for these kind of galaxies. In
Figure~5b, we observe the same behavior: some LINERs have sulphur
line ratios similar to SFGs, while some SCG SFGs show unusually
strong sulphur lines.  The separation of LINER and SFGs is clearer
in Figure~5c, reflecting the fact that we used preferentially this
diagram to classify our galaxies.

From the analysis of 200 luminous infrared galaxies, Veilleux et
al. (1995) have shown that in 81\% of the cases the diagnostic
diagram [\ion{O}{3}]$\lambda5007$/H$\beta$ vs
[\ion{N}{2}]$\lambda6584$/H$\alpha$ is sufficient to classify
without ambiguity the activity types of the galaxies.  Using two
other diagnostic diagrams, namely
[\ion{O}{3}]$\lambda5007$/H$\beta$ vs
[\ion{S}{2}]($\lambda6717+\lambda6731$)/H$\alpha$ and
[\ion{O}{3}]$\lambda5007$/H$\beta$ vs
[\ion{O}{1}]$\lambda6300$/H$\alpha$, they found only 19\% of
galaxies with an ``ambiguous'' classification, that is, they show
different activity types in different diagnostic diagrams. The
same result was recently found by Contini et al. (1998) using a
sample of 105 barred SFGs, and in general, for most emission--line
galaxies ambiguous classification seems rare (V\'eron et al. 1997;
Coziol et al. 1999).  In their sample of 418 ``normal'' galaxies
more luminous than B$_T \leq 12.5$, Ho et al. (1997a, 1997b) found
a much larger number of galaxies with ambiguous classification
using standard diagnostic diagrams. These star--forming galaxies
usually have anomalous strengths of low--ionization lines. They
associated this behavior to the \ion{H}{2} Nucleus Galaxies
phenomenon (Kennicutt et al. 1989; Coziol 1996). In Figure~5, we
show that most of the SCG SFGs do occupy in these diagrams the
\ion{H}{2} Nucleus Galaxies regions (the gray areas in Figure~5b
and 5c). It seems, therefore, that the ambiguity of the galaxies
in our sample can be explained simply if they are, in fact,
\ion{H}{2} Nucleus Galaxies.

In Coziol (1996) it was shown that the star formation in the
\ion{H}{2} Nucleus Galaxies is generally at a lower level than in
starburst galaxies. Therefore, we can use a star formation
indicator like EW(H$\alpha$) to sheck if the galaxies in our
sample are \ion{H}{2} Nucleus Galaxies.  In Figure 6, we show the
distribution of EW(H$\alpha$) for all the emission--line galaxies
in our sample. In this figure, we do not include the galaxy SCG
0105-1744Ac, which, with an H$\alpha$ equivalent width of 290\AA,
is an obvious starburst galaxy. The rest of the SCG SFGs, on the
other hand, have relatively narrow equivalent width. These values
are typical of \ion{H}{2} Nucleus Galaxies (Coziol 1996),
supporting, therefore, our interpretation.

As for the nature of \ion{H}{2} Nucleus Galaxies and the origin of
their peculiar spectral characteristics, Kennicutt et al. (1989)
and Gon\c{c}alvez et al. (1999) suggested that they could be a
sort of intermediate type galaxies, where star formation and AGN
activity coexist in their nucleus. To test this hypothesis, we
have plotted in Figure~7 the diagnostic diagram
[\ion{O}{2}]$\lambda3727$/[\ion{O}{3}]$\lambda5007$ vs.
[\ion{O}{1}]$\lambda6300$/[\ion{O}{3}]$\lambda5007$ for the SCG
galaxies.  This diagram was originally used by Heckman (1980) to
define the LINER type.  In this figure, the horizontal and
vertical lines indicate the separations, as defined by Heckman,
between Seyfert 2 and LINERs.  The areas in gray define
schematically the locus occupied by Seyfert 2 (SY2), LINERS and
transition type galaxies (T). For comparison, we also included in
this figure starburst galaxies as observed by Veilleux et al.
(1995) as well as data for normal \ion{H}{2} regions as observed
by van Zee et al. (1998). The differences between these different
samples are interesting. The position of the few SCG that we can
put on this plot is quite distant from the locus of normal HII
regions and starburst. They all fall in the LINER region or in the
transition region between Seyfert 2 and LINERs, consistent with
the interpretation of Kennicutt et al. 1989 and Gon\c{c}alvez et
al. 1999 for the \ion{H}{2} Nucleus Galaxies. It is important to
emphasize that this result does not depend neither on the way our
spectra were corrected for the dilution effect (the forbidden
lines are not affected by the underlying stellar populations) nor
on the amount of reddening present. It is therefore safe to assert
that in the SCGs sample most of the galaxies do not show strong
star forming activity, neither the typical behavior of normal HII
regions, but possibly a mixture of star formation and AGN
activity, as indicated by Figure 7.

Regarding the location of the activity in the AGN, LLAGN, and HII
Nucleus Galaxy candidates, it is important to mention that our
spectroscopic observations cover the most luminous part of the
galaxies and are centered on the nucleus suggesting that the
peculiar phenomena we observe is located in these regions of the
galaxies. However, we lack the spatial resolution to say that this
is strickly a nucleus phenomenon. In Coziol et al. (1998a), it was
verified that after a template subtraction, the line ratios and
equivalent widths of LLAGN are consistent with those of LINERs and
Seyfert galaxies. This phenomenon is also frequently observed in
the nucleus of normal elliptical and S0 galaxies (Rubin et al.
1991). Therefore, for these objects, it seems that we are dealing
with a nuclear phenomenon.

But for the star forming galaxies the situation is still unclear.
These systems seem to show an excess of excitation suggesting some
kind of transition phase between LINERs and starburst. In these
cases, we cannot conclude that this extra excitation is
concentrated (or have its origin) in the nucleus. We have already
found similar ambiguous situations in HCG 16 (see de Carvalho \&
Coziol 1999). More observations will be necessary to settle this
important question.

\subsection{Comparison with the HCGs sample}

In this section, we proceed to verify if our classification of
activity types in the SCGs sample is consistent with what was
observed previously in the HCGs sample. We also verify if we
recognize the same luminosity--activity and morphology--activity
found in these compact groups (Coziol et al. 1998a).

In Table~6, we compare the frequencies of the different activity
types as observed in the HCGs and in the SCGs. For this analysis,
we added the few (5\%) EMLGs in Table~1 to the SFGs. For the HCGs,
we give both the frequencies for the groups and the core.
According to our interpretation of the core and halo structure in
CGs (Ribeiro et al. 1998; Coziol et al. 1998a), and considering
the criteria to select the SCGs, our results should be similar
mostly to those of the core of the HCGs. In Table~6, we can see
that the distribution of activity types observed in the SCGs is
comparable to the one found in the HCGs cores. In particular, low
luminosity AGNs and AGNs are, as for the HCG cores, the most
numerous activity types found in SCGs (41\% of the cases).  We
note however an interesting difference with the HCG sample: the
number of SFGs is larger while the number of non--emission
galaxies is lower.

We look now for the presence of a luminosity--activity relation in
SCGs.  According to the relation found for the HCGs, we expect to
see the low luminosity AGNs and AGNs preferentially located in the
most luminous galaxies in the groups, while the
non--emission--line galaxies and the SFGs should share a common
distribution among the less luminous galaxies.  Figure~8 shows the
luminosity distribution for the SCGs galaxies.  Comparing
Figure~8a with Figure~8b, we do see a clear tendency for the low
luminosity AGNs and AGNs to be located in the more luminous
galaxies of the groups.  This tendency is reinforced when we
consider in Figures 8c and 8d the groups with only 4 or more
members.  This is similar to the luminosity segregation previously
observed in HCGs.

A Kolmogorov-Smirnov test was performed on the magnitude
distributions of the different types. The results are presented in
Table~7. It gives the probabilities that the different types of
galaxies come from the same distribution. These results support
the interpretation that AGN and LLAGN in SCGs share the same
magnitude distribution. However, they do not support the
interpretation that the non emission line galaxies and the SFGs
come from the same distribution. The SFGs seem clearly different
from the other types of galaxies.

Finally, we show in Figure~9 the distribution of the morphologies
in the SCGs.  For the HCGs a morphology--activity relation links
the non--emission--line galaxies, low luminosity AGNs and AGNs to
the most early--type galaxies and the SFGs to the late--type ones.
This result holds independently from the core--halo structure
found in CGs.  This trend is confirmed also in our sample. In
particular we confirm that the SFGs are located mostly in
late--type spiral galaxies.  We also observe another difference
between the SCGs and HCGs:  a larger number of low luminosity AGNs
in the SCGs are in early--type spirals (Sa and Sab).

From the above comparisons, we conclude that we observe in the
SCGs the same characteristic activity trends as in the HCGs. The
SCGs are dominated by AGNS and galaxies without emission, and the
AGNs are located in the most luminous and early--type galaxies of
the group. We confirm also that the SFGs are located mostly in
late--type spiral galaxies.

But we also observe some interesting differences between the SCGs
and the HCGs. The SCGs contain more SFGs and less galaxies without
emission than the HCGs. The SCGs contain also more LLAGNs in
spirals than the HCGs. These differences may be interpreted in
different ways.

If CGs are real structures evolving by interaction, we should then
expect the less evolved structures to be those richer in spirals
and SFGs. The differences between the SCGs and HCGs could have,
therefore, a physical origin: the SCGs contain less evolved
systems than the HCGs. The difference between the two samples
could be explained by the biases affecting the HCGs sample. Being
selected in an automated way, the SCGs probe a wider range of
physical properties than the HCGs, whose selection biases favor
groups with extreme properties, denser and more evolved, because
more easily detectable being more prominent on the plates.

Alternatively, the SCGs may also contain more SFGs than the HCGs
because they are more severely affected by projection effects. The
fact that most of the SFGs are actually \ion{H}{2} Nucleus
Galaxies and that these galaxies are very common in loose groups
(Ho et al. 1997a) may support this alternative.

To test the different hypothesis, we present in the last column of
Table~6, the activity frequency in the SCGs with four or more
members. We see a significant increase of galaxies without
emission accompanied by a comparable decrease of the number of
SFGs.  The larger number of SFGs in the SCGs seems, therefore, to
be due to their preponderance in the subset of groups with only 3
concordant members. It is difficult, however, to discriminate if
this is due to an evolution or to a projection effect, as the
probability of having a projection group increases when the number
of members decreases. To help disentangle the above problem, we
can look at the activity--type association in the SCGs.  If
compact groups are projection effects within loose groups, then
the majority of those with three members should be mostly composed
of SFGs. But, out of the 16 triplets in our sample, only two
contain three SFGs. Considering our whole sample, only 22\%
contain a majority of SFGs (compared to 44\% in the triplets). In
general, the groups with SFGs also include LLAGNs (47\%) or
galaxies without emission (55\%).

Although, the above results do not completely exclude projections,
they suggest that this is probably not the only reason for the
differences observed between the SCGs and the HCGs. The SCGs
sample, therefore, may also includes a greater fraction of groups
which are relatively less evolved than in the HCGs.

As a last point, there are probably very simple reasons why we did
not discover the predominance of \ion{H}{2} Nuclear Galaxies among
the SFGs in our previous survey of 17 HCGs. First, our previous
survey did not include emission lines like
[\ion{O}{2}]$\lambda3727$ or emission line ratios like
[\ion{S}{2}]/H$\alpha$. Our sample was also smaller and
contaminated by SFGs from the halo of the groups, where this
phenomenon may be less pronounced. Our sample contain also the
galaxies in HCG 16, which are unusually active in star formation
(Ribeiro et al.  1996; de Carvalho \& Coziol 1999). However, a
careful examination of our previous observations in HCGs does
reveal many peculiar characteristics (Coziol et al.  1998b). All
the line ratios of the HCGs SFGs are consistent with those of
\ion{H}{2} Nucleus Galaxies.  Many of these galaxies also have
relatively strong [\ion{O}{1}]$\lambda6300$ lines as compared to
starburst galaxies and some show
[\ion{O}{1}]$\lambda6300$/H$\alpha$ suggesting an ambiguous nature
(starburst--LINER). Finally, in general in the HCGs, the H$\alpha$
FWHM of the SFGs are unusually large for starburst galaxies
(Coziol et al.  1998b). Even in HCG 16, where star formation is at
a higher level than in any other HCGs, we see some evidence of an
ambiguous nature in some galaxies: HCG16-1 is a LINER with a
circumnuclear starburst, while HCG16-5 have a mixture of AGN and
star formation in its nucleus (de Carvalho \& Coziol 1999).
Selection biases favoring more evolved groups in the HCGs sample
would also have rendered evidence of the predominance of
\ion{H}{2} Nuclear Galaxies less obvious in our first sample.

\section{Summary and conclusion}

We have presented the results of the classification of the level
of activity of 193 galaxies in a new sample of 49 CGs of galaxies
observed in the southern hemisphere. This analysis confirms the
results previously obtained from a smaller sample of 17 HCGs.

\begin{itemize}

\item We show that AGN (including low luminosity AGN) is the
most frequent (41\%) activity type encountered in CGs. We found
however more SFGs and less non--emission--line galaxies in the
SCGs than in the HCGs. We suggest that these differences could be
explained by the visual biases affecting the HCG sample. The SCGs
sample, filling completely the parameter space allowed by the
criteria used for the search, contain groups with a broader range
of properties than the HCGs sample, and therefore groups whose
properties (\eg spatial density) are less extreme than those of
HCGs.

\item We confirm the luminosity--activity and morphology--activity
relation previously observed in HCGs: 1) low luminosity AGNs and
AGNs are preferentially located in the most luminous galaxies in
the groups, while non--emission--line galaxies and SFGs share a
common distribution among the less luminous galaxies; 2) the
non--emission--line galaxies, the low luminosity AGNs and AGNs are
more common in early--type galaxies while SFGs are preferentially
located in late--type spirals.

\end{itemize}

We also observed some new phenomena:

\begin{itemize}

\item We find a strong variation of the dominant activity type with
the increase of galaxies members in the group: the number of
non--emission--line galaxies increases while the number of SFGs
decreases. But what truly characterizes CGs seems to be the
association of one or more SFGs with more evolved or active
galaxies (low luminosity AGNs, LINERs or Seyfert 2). It is quite
rare to find a group composed only of SFGs, which suggests that
what we are looking at is really a physical entity, and not just a
projection effect within loose groups.

\item The star--forming galaxies in CGs are mostly \ion{H}{2} Nucleus
Galaxies. As these galaxies are characterized by having low star
formation activity (Coziol 1996), we conclude that the star
formation level in CGs is generally low. This conclusion is
consistent with the recent results on star formation obtained by
Severgnini et al. (1999) based on an H$\alpha + [{\rm H II}]$
imaging survey in 31 HCGs. Comparing the distribution in H$\alpha
+$ [\ion{N}{2}] EW measured by these last authors with the
distribution presented in Coziol (1996) for the \ion{H}{2} Nucleus
Galaxies (multiplying by 0.66 to 0.5 to take into account the
unusual strengths of the nitrogen lines in these galaxies), one
can see that they are fully consistent.

\item We also see in the \ion{H}{2} Nucleus
Galaxies a clear evidence for non stellar activity.  If the
transition nature of these galaxies is taken as a sign of AGN
activity, then CGs would be exceptionally rich in AGNs. More than
70\% of the galaxies in the SCGs would have an AGN nature (adding
the SFGs to the AGNs and LLAGNs). The AGN nature of these
galaxies, however, must be confirmed, as alternatives, like shocks
produced by supernovae or excess of excited gas in the disk could
also be valid.

\end{itemize}

To understand which mechanisms in CGs produce the phenomena
described above is not an easy task.  As was already observed
(Coziol 1998a; 1998b), we cannot relate directly the high density
in compact groups of galaxies to an increase (or decrease) of star
formation or to an increase of nuclear activity.  What we observe
instead is a sort of density--morphology--activity relation.  The
reason for this multiple relation may have something to do with
the history of galaxies in CGs and the mechanisms by which these
systems form.

The dynamical analysis of Ribeiro et al.  (1998) and Barton et al.
(1998) reveal that most of the compact groups are some sort of
substructure of larger structures.  These substructures are not
all at the same dynamical evolutionary stage.  Some may already be
in dynamical equilibrium while others are still replenished in
spiral galaxies, which fall from the larger structure where they
first formed (Diaferio et al.  1994; Governato et al.  1996). This
is specially true for SCGs, where the spread in physical
properties is bound to be larger than that of the HCGs.

According to the scenario of sub--clustering in loose groups
(Diaferio et al. 1994), during the expansion and collapse of
gravitating systems subclumps form and merge in a hierarchical
manner. The subclumps are often bound and can be identified with
compact groups. In their models, Diaferio et al. 1994 have found
that most of the subclumps contain 2 or 3 galaxies. The frequency
for groups with 4 and 5 galaxies are, at maximum, 34\% and 12\%,
respectively. These numbers are comparable to what we find: 46\%
and 13\% SCGs have 4 and 5 members, respectively. However, to be
consistent with the simulations done by Diaferio et al. (1994),
the loose groups where SCG would have formed would need to be
richer in galaxies. Loose groups with 15 and 30 galaxies are
necessary to reproduce our observations. In their study of poor
groups with diffuse X-ray, Zabludoff \& Mulchaey (1998) find a
number of galaxies between 20 and 50. Barton et al. (1998), on the
other hand, have found more heterogeneous environments, the number
of galaxies in their sample varying between 5 members up to the
population of rich clusters of galaxies. It is not clear which
results are representative of the environment of CGs in general,
but it seems that the Diaferio et al. scenario of sub--clustering
could a good working hypothesis.

However, there could be a serious problem with the sub--clustering
scenario, in the relatively short lifetime it predicts for these
systems. After $\sim 1$Gyr the galaxies should all merge. In the
Diaferio et al. simulations, CGs are currently forming and,
consequently, are not sufficiently evolved to contain a large
number of merger remnants. This in contradiction with the advance
level of evolution of the galaxies usually found in the core of
CGs (Coziol et al. 1998a; 1998b). This fast merger scenario is
also in contradiction with our observation of the variation of
activity status of the galaxies with the number of group members:
the groups which have a high number of galaxies are also the ones
which are more evolved and less active. Furthermore, while the
Diaferio et al. model correctly predicts that the more massive
galaxies form the CGs, it does not explain how these galaxies were
allowed to grow in mass. It seems reasonable to assume that
mergers must have taken place sometime in the evolution of these
structures to produce both the more massive and early--type
galaxies necessary to form the core of the groups. But, as we have
emphasized before, signs of mergers are generally scarce, and in
the more evolved galaxies in the groups star formation activity
has probably ceased more than 2 Gyrs ago (Caon et al. 1994; Coziol
et al. 1998b). It seems, therefore, that another parameter is
missing in our present history of the formation of these
structures which would allow galaxies to survive merging.

The replenishment scenario (Governato et al.  1996) offers one
simple solution to the fast merger problem. According to this
scenario, if CGs are so conspicious it is because they are
continuously replenished by galaxies falling into their core. Now,
if we assume that the infalling galaxies are preferentially
late--type SFGs, we would then expect to find a high number of
these galaxies in each group if replenishment is important.
Moreover, these infalling galaxies should be associated
preferentially with galaxies with more evolved characteristics. We
have verified that late--type SFGs are indeed present in 80\% of
the groups. We have also shown that they are preferentially
associated to other activity types, in particular LLAGNs and
no--emission line galaxies which are the more evolved galaxies in
CGs. This supports the idea that replenishment is an important
aspect of the compact group phenomenon (Governato et al. 1996).

According to the replenishment scenario, the SFGs would be ``now''
falling into the core of the groups.  By doing so, they will
interact with the group as a whole, but possibly also with
individual galaxies. For CGs, the models describing these
interactions should explain both the small amount of star
formation observed in these galaxies and the activation of
non--thermal process.

Simulations of low--velocity galaxy--galaxy interaction have shown
that the majority of galaxy--galaxy encounters lead to
normal--star forming properties and that the triggering of very
active starbursts are rare events (Mihos, Richstone \& Bothun
1991; 1992).  In some of the closest encounters star formation
could even be inhibited as a result of mass loss from the disk (by
tidal tails) or perturbations that increase the disk gas scale
height (Mihos, Richstone \& Bothun 1991). Only mergers, seem to be
able to increase significantly the star formation rate in
interacting galaxies (Mihos, Richstone \& Bothun 1992). At first,
these results seem consistent with the low star formation observed
in CGs. The small number of starburst galaxies in these systems
could also be the result of rare mergers, as it is clearly
suggested for some galaxies in HCG 16.  According to the
simulations, however, low--velocity galaxy--galaxy interactions
would systematically produce morphological evidences (Howard et
al.  1993).  Although we do see peculiarities in some of the SFGs,
morphological distortions are not always present. According to
simulations, it is also very difficult to induce star formation in
the nucleus of galaxies by galaxy--galaxy interaction. Even more
so to move the gas toward the center to feed a possible AGN.

Considering that gravitational forces are additive, it is probably
more realistic to assume that a galaxy which is falling into the
core of the group will first interact with the potential of the
group as a whole. In the literature, one can find many models
explaining how environmental effects may change the evolution of
galaxies falling into rich clusters. Some of the mechanisms
proposed in these models may also apply to galaxies in CGs. In
some aspects, CGs do resemble rich clusters of galaxies. The
density of galaxies in the core of CGs, for instance, is
comparable to those in clusters. On other aspects, however, CGs
are quite distincts. The galaxies in CGs, in particular, have
smaller velocity--dispersions and there is much less galaxies in
number. The differences should limit the efficiencies of the
possible mechanisms in rich clusters and produce some notable
differences.

The effect of tidal interaction between cluster gravitational
potential and a disk galaxy was studied by Merritt (1982; 1983),
Byrd \& Valtonen 1990, Henriksen \& Byrd (1996) and more recently
Fujita (1998).  According to Merritt (1982; 1983) a galaxy falling
into a rich cluster will have its outer parts rapidly stripped
away by the tidal field. But, before tidal stripping becomes
effective, the perturbation produced by the tidal field can also
trigger an inflow of gas toward the center of the galaxies which
may either feed an AGN (Byrd \& Valtonen 1990) or start a burst of
star formation (Henriksen \& Byrd 1996; Fujita 1998).  Using a
simple point--mass model, Byrd \& Valtonen (1990) have shown that
infall of matter happens long before tidal stripping in the disk
becomes effective. Therefore, gas rich spiral falling into the
group will become active long before reaching the center.

The consequences of this model are interesting. Among other
things, it predicts that CGs should have a halo of SFGs or even
starburst galaxies around them. When the spiral galaxies enter the
core of the group, tidal stripping becomes effective and the disks
loose part of their gas (Hutchmeieir 1997). Tidal stripping will
not touch the molecular gas which have formed in the center of the
galaxy (Verdes--Montenegro 1997; Leon et al. 1998), but it will
stop the infall of gas and decrease the star formation in the
disk. The galaxies may then resemble an \ion{H}{2} Nucleus
Galaxies, with a small residual of star formation and non thermal
activity in their nucleus.

If the source of the non thermal activity is the feeding of an AGN
in the center of the galaxies, the question which naturally arise
is why is the formation of Seyfert 1 inhibited in CGs?  One reason
may be because the infall of gas is followed automatically by star
formation. The AGN would then have to compete with star formation
for the gas (de Carvalho \& Coziol 1999). When star formation
decreases, a low luminosity AGN then becomes apparent.

Tidal triggering added to tidal gas stripping will combine their
effects to transform late type spirals into early--type ones.
Triggered star formation in the nucleus of the galaxies increase
their bulge in mass and light, while gas stripping makes them
loose their disk. Although such a scenario seems viable, an
important caveat is the fact that the total mass involved in CGs
are small compared to those of rich clusters, making the
efficiency of these transformations quite low, which could explain
why a significant fraction of CGs are still dominated by spiral
galaxies.

Other mechanisms could be important in rich cluster of galaxies
but probably do not take place in compact groups. These are ram
pressure gas stripping and galaxy harassment. Ram pressure gas
stripping should be less efficient in CGs, since it increase as
$\sim L_x^{1/2} v_cg^2$. Although CGs may exhibit X-ray
luminosities similar to clusters (Ponman et al. 1996), the
velocity dispersions are much smaller. Galaxy harassment (Moore,
Lake \& Katz 1998; Lake. Katz \& Moore 1998), on the other hand,
needs frequent collisions at high speeds, which is difficult to
achieve in CGs, where galaxies have lower velocity and there are
also much fewer galaxies to interact with. Comparison of the
evolution of CGs galaxies with those in rich clusters may allow to
distinguish the specific signatures of these two particular
mechanisms in rich cluster of galaxies.

\acknowledgments

R. C. would like to thank the Osservatorio di Brera and the GNA
for the funding of his visit to the Observatorio di Brera, Milano,
during which this paper was written.  He would like also to thank
the direction and personnel of the Osservatorio di Brera for their
hospitality. The authors also acknowledge the anonymous referee
for his comments and suggestions which have contributed to improve
the article.

\clearpage
\begin{deluxetable}{ccllccccll}
\footnotesize \tablecaption{Compact Groups\label{tbl-1}}
\tablewidth{0pt} \tablehead{ \colhead{Gal. id.}&\colhead{$B_j$} &
\colhead{Morph.}     &\colhead{Activity} &&& \colhead{Gal.
id.}&\colhead{$B_j$} & \colhead{Morph.}     &\colhead{Activity}
\\ \colhead{ }       &\colhead{} & \colhead{ }
&\colhead{Type} &&& \colhead{ }       &\colhead{} & \colhead{ }
&\colhead{Type} } \startdata 0004-5044: &        &        & &&&
0017-5129:&&    &        \\
         A &  14.22 & S0     & no em. &&& A &  14.29 & Sb & SFG    \\
         B &  15.60 & Sa     & LLAGN  &&& C &  15.53 & S0 & LLAGN? \\
         C &  15.87 & S0     & LLAGN  &&& D &  16.03 & Sb & SFG    \\
         D &  16.01 & Sc     & SFG    &&&   &        &    &        \\
0018-4854\tablenotemark{n}:&&& &&&0031-2143\tablenotemark{n}:&
(HCG 004)&   & \\
         A &  13.91 & Sa pec & LINER  &&& A &  14.02 &  Sb   & SFG    \\
         B &  14.34 & Sb pec & LINER  &&& C &  16.38 &  E    & no em. \\
         C &  14.48 & Ir     & SFG    &&& D &  16.47 &  E    & SFG    \\
         D &  14.97 & Sb pec & LINER  &&&   &        &       &        \\
0034-2250: &        &         &       &&&0035-3357: &&         &
\\
         A &  14.07 &  Sb     & LLAGN &&& A &  13.92 &  pec    & LINER \\
         B &  14.59 &  Sc     & SFG   &&& B &  15.82 &  Sc     & no em.\\
         C &  15.25 &  Sa     & SFG   &&& C &  16.13 &  Sc     & LINER \\
           &        &         &       &&& D &  16.28 &  Sb     & SFG   \\
0040-2350:                  &        &         & &&&0054-5312:& &
&        \\
         A\tablenotemark{n} &  14.21 &  Sa  pec & LLAGN     &&& A &  13.27 &  E      & no em. \\
         Ac\tablenotemark{n}&  14.21 &  E  pec & Sy2       &&& B &  13.41 &  E      & no em. \\
         B                  &  14.43 &  Sb     & LINER     &&& C &  13.97 &  Sc     & LLAGN  \\
         C                  &  15.62 &  Sc     & SFG       &&& D &  14.07 &  S0     & LLAGN  \\
                            &        &         &           &&& E &\nodata & \nodata & LLAGN  \\
                            &        &         &           &&& F &\nodata & \nodata & no em. \\
0100-2208: &        &         &        &&& 0102-4714:&&     &
\\
         A &  14.45 & E/S0 & no em. &&& A &  14.38 &  E  & no em.  \\
         C &  16.56 & \nodata   & LLAGN? &&& B &  15.63 &  Sa & SFG     \\
         D &  16.86 & \nodata   & no em. &&& C &  15.66 &  E  & LLAGN   \\
           &        &         &        &&& D &  16.63 &  Sa & LLAGN?  \\
0105-1744\tablenotemark{n}: &        &      &      &&&
0106-4722:&&     &        \\
         A                  &  14.25 &  pec & LINER&&& A &  14.40 &  S0 & LLAGN? \\
         Ac                 &  14.25 &  pec & SFG  &&& B &  14.82 &  Sa & LLAGN? \\
         B                  &  15.06 &  Sc  & SFG  &&& C &  14.96 &  Sa & LLAGN  \\
         D                  &  17.20 &  ?   & SFG  &&& D &  15.84 &  Sa & LLAGN  \\
0111-3203: &        &        &        &&& 0116-4439:         & &
&       \\
         A &  12.58 &  E     & LLAGN? &&& A\tablenotemark{n} &  13.73 &  pec   & LINER \\
         B &  13.68 &  Sa    & LLAGN  &&& B                  &  15.38 &  Sc    & LINER \\
         C &  15.17 &  E     & LLAGN  &&& D                  &  16.29 &  Sb    & LINER \\
         D &  15.38 &  Sa    & LLAGN  &&&                    &        &        &       \\
0118-3624: &         &      &      &&& 0122-3819:& &        &
\\
         A &  14.44  & Sb   & SFG  &&& A &  14.23  & S0      &  no em. \\
         B &  16.48  & E    & Sy2  &&& B &  14.23  & E      &  LLAGN  \\
         C &  16.69  & pec  & EMLG &&& C &  14.54  & Sb     &  SFG    \\
           &         &      &      &&& D &  15.54  & S0     &  no em. \\
0146-2721: &         &    &      &&& 0146-4855:& &     &       \\
         A &  14.09  & Sc & SFG  &&& A &  12.97  & Sc  & LLAGN \\
         C &  15.78  & Sc & LINER&&& B &  14.85  & ?   & LINER \\
         D &  16.81  & S0 & LLAGN&&& C &  15.30  & Ir  & SFG   \\
         E &  16.91  & S? & SFG  &&&   &         &     &       \\
0147-3506: &         &    &        &&& 0156-5629:& &        &
\\
         A &  14.30  & Sa & LLAGN? &&& A &  14.45  & E      & LLAGN  \\
         B &  14.60  & Sc & SFG    &&& B &  14.74  & S0     & LINER  \\
         C &  16.70  & Sb & EMLG   &&& C &  16.85  & E      & no em.?\\
           &         &    &        &&& D &  17.11  & E      & LINER  \\
0227-4312: &         &        &        &&& 0242-1750: &(HCG 21)&&
\\
         A &  12.93  & S0      & no em. &&& A &  13.94 &  E      & LLAGN  \\
         B &  13.70  & Sb     & EMLG   &&& B &  14.12 &  Sa     & LLAGN? \\
         C &  13.99  & Sb     & no em. &&& C &  14.14 &  Sc     & EMLG   \\
         D &  14.55  & Sb     & LLAGN  &&&   &        &         &        \\
         E &  14.79  & Sa     & SFG    &&&   &        &         &        \\
         F &  14.86  & Sa     & no em. &&&   &        &         &        \\
0328-4807: &        &      &      &&& 0414-5559:&&         &
\\
         A &  13.87 &  Sb  & SFG  &&& A &  10.86 &  E:     & LLAGN \\
         B &  14.58 &  Sa  & LINER&&& B &  11.20 &  E:     & no em.\\
         C &  15.23 &  Sc  & SFG  &&& C &  12.25 &  Sab:   & SFG   \\
           &        &      &      &&& D &  13.70 &  Sc:    & SFG   \\
0421-5624: &        &      &      &&& 0426-4249: & &&        \\
         A &  14.39   & Sa & no em. &&& A &  14.20 &  E pec    & LLAGN  \\
         Ac& \nodata  & E  & no em. &&& Ac&  14.20 & \nodata& no em. \\
         B &  14.65   & Sb    & Sy1?   &&& B &  14.30 &  S0    & no em. \\
         F &\nodata   & ?     & SFG    &&& C &  14.45 &  Sc    & SFG    \\
0427-4802: &        &        &        &&& 0435-5131:&&     &
\\
         A &  13.33 &  Pec & Sy2    &&& A &  14.13 &  E  & no em. \\
         B &  13.94 &  Sc    & SFG    &&& B &  16.43 &  Sc & SFG    \\
         C &  14.02 &  S0     & no em. &&& C &  16.62 &  S0 & LINER  \\
           &        &        &        &&& D &  16.68 &  Sa & LLAGN? \\
0456-2141: &        &     &       &&& 0533-5057:&&     &     \\
         A &  14.45 &  Sc & LINER &&& A &  14.37 &  Sc & SFG \\
         B &  14.63 &  Sb & LLAGN &&& C &  16.48 &  S? & SFG \\
         C &  15.63 &  Sc & SFG   &&& D &  17.35 &   ? & SFG \\
         D &  15.76 &  S0 & LLAGN?&&&   &        &     &     \\
0608-4734: &        &     &        &&& 2113-4235:&&       &
\\
         A &  13.97 &  E  & no em. &&& A &  13.01 &  E    & LLAGN \\
         B &  15.81 &  E  & LINER  &&& B &  13.81 &  Sc   & SFG   \\
         C &  16.05 &  S0  & no em. &&& C &  13.86 &  Sa   & no em.\\
         D &  16.60 &  S0 & Sy2    &&& D &  14.72 &  S0   & no em.\\
         E &  16.76 &  S0 & no em. &&&   &        &       &       \\
         F &\nodata &  E  & no em. &&&   &        &       &       \\
2114-2302: &        &      &       &&& 2122-4245:&&             &
\\
         A &  14.20 &  Sc  & no em.&&& A &  13.57 &  Sb         & LLAGN  \\
         B &  14.21 &  Sc  & LINER &&& B &  13.75 &  E          & no em. \\
         C &  14.75 &  Sa  & LINER &&& C &  14.65 &  Sa          & no em. \\
         D &  15.31 &  Sb  & SFG   &&& D &  15.08 &  Sb         & SFG    \\
           &        &      &       &&& F &\nodata &  S0         & no em. \\
2123-2259: &        &       &      &&& l2123-6105:& &       &
\\
         A &  14.24 &  SBc  & LINER&&& A &  14.20   &  Sb   & LLAGN  \\
         B &  14.60 &  Sc   & SFG  &&& B &  14.53   &  Sa   & no em. \\
         C &  16.62 &  Sa   & SFG  &&& C &  15.40   &  Sb   & no em. \\
           &        &       &      &&& E &  16.70   &  S?   & EMLG   \\
2124-2314:  &        &          &       &&& 2128-4614:&& &
\\
          A &  14.43 &  Sb      & Sy2   &&& A &  14.35 &  Sb      & LLAGN \\
          B &  15.58 &  E       & no em.&&& B &  15.10 &  E       & SFG   \\
          C &  17.06 &  S?      & SFG   &&& C &  16.20 &  Sc      & EMLG  \\
          D &  17.41 &  S?      & LINER &&& D &  16.51 &  Sa      & SFG   \\
            &        &          &       &&& E &  16.69 &  S0      & EMLG  \\
2146-5935:  &        &     &       &&& 2200-2048:&&     &
\\
          A &  14.42 &  E  & LLAGN &&& A &  11.93 &  Sc & LLAGN  \\
          B &  15.86 &  S0 & SFG   &&& B &  13.51 &  S0 & no em. \\
          C &  15.94 &  Sb & SFG   &&& C &  14.02 &  Sc & SFG    \\
            &        &     &       &&& D &  14.13 &  S0 & LLAGN  \\
2211-4616:  &        &        &       &&& 2213-2743:&&       &
\\
          A &  12.70 &  Sa    & LINER &&& A &  14.12 &  S0:  & no em.\\
          B &  12.70 &  Sa    & LLAGN &&& B &  15.19 &  Sb?: & SFG   \\
          C &  12.95 &  Sb    & SFG   &&& C &  16.21 &  E?:  & LLAGN?\\
          D &  13.79 &  S0    & SFG   &&& D &  16.85 &  E?:  & no em.\\
          E &  14.44 &  Ir    & LINER &&&   &        &       &       \\
          F &\nodata & ?      & SFG   &&&   &        &       &       \\
2226-3546:  &        &         &       &&& 2226-6111:&&      &
\\
          A &  14.16 &  E      & no em.&&& A &  13.62 &  E   & no em.\\
          B &  16.46 &  Sa     & no em.&&& B &  15.50 &  Sa  & SFG   \\
          C &  16.56 &  Sb     & EMLG  &&& C &  15.78 &  Sc  & SFG   \\
          D &  16.92 &  Sb     & no em.&&& E &  16.00 &  Sa  & no em.\\
          E &\nodata &  ?      & EMLG  &&& F &  15.77 &  S0  & no em.\\
2315-4241:  &(Grus Group)  &  &      &&& 2316-2259:&&        &
\\
          A &  11.34 &  Sb    & Sy2  &&& A &  14.30 & SBc    & LINER\\
          B &  11.47 &  Sb    & LINER&&& B &  14.68 & pec    & SFG  \\
          C &  11.71 &  Sc    & LINER&&& C &  16.64 & Sc     & SFG  \\
          D &  12.35 &  Sb    & LINER&&&   &        &        &      \\
2351-4015:  &        &        &        &&& 2353-6101:&&         &
\\
          A &  14.46 & E      & LLAGN? &&& A &  13.96 &  Sb     & SFG   \\
          Ac&\nodata & S0     & no em. &&& B &  14.03 &  Sc pec & LINER \\
          B &  16.79 & S0     & no em. &&& C &  14.91 &  S0     & no em.\\
          C &  17.04 & S0     & no em. &&&   &        &         &       \\
          D &  17.35 & E      & no em. &&&   &        &         &       \\
2358-4339:  &        &     &         &&&   &        &       &
\\
          A &  13.79 &  E  & no em.  &&&   &        &       &       \\
          B &  15.92 &  S0 & LLAGN   &&&   &        &       &       \\
          C &  16.36 &  S? & no em.  &&&   &        &       &       \\
          D &  16.38 &  S0 & no em.? &&&   &        &       &       \\
\enddata
\tablecomments{ 0031-2143: This group corresponds to HCG 004. The
spectral classifications were taken from Coziol et al. (1998a).
The morphologies are as found in NED. 0018-4854: Obvious case of
merger. Could be a young group in formation like HCG 16;
0040-2350A \& 0040-2350Ac: these two galaxies seem to be in a
merger phase. Strangely, only the smaller mass one has a Seyfert 2
in its nucleus; 0105-1744: Another obvious case of merger and
possible young group in formation; 0116-4439A: Obvious case of two
early--types spirals in merger phase; 2124-2314A: Giant elliptical
with shell or dust ring and a Seyfert 2 in its nucleus; 2315-4241:
The Grus Group. A very well known group of galaxies. Contrary to
other cases of ``young'' or HCG 16 like group, the galaxies here
are all giant evolved spirals.}
\end{deluxetable}

\newpage
\begin{deluxetable}{lcccccc}
\footnotesize \tablecaption{Spectral line ratios in SCG's
emission-line galaxies. \label{tbl-2}} \tablewidth{0pt}
\tablehead{ \colhead{SCG \#} &\colhead{$\frac{{\rm [NII]}}{{\rm
H}\alpha}$} &\colhead{$\frac{{\rm [OIII]}}{{\rm H}\beta}$} &
\colhead{$\frac{{\rm [SII]}}{{\rm H}\alpha}$}
&\colhead{$\frac{{\rm[OI]}}{{\rm H}\alpha}$}  &
\colhead{$\frac{{\rm [OI]}}{{\rm [OIII]}}$}
&\colhead{$\frac{{\rm [OII]}}{{\rm [OIII]}}$} } \startdata
0004-5044D      &   0.44$\pm$0.04 &      \nodata    &
0.43$\pm$0.07 &   0.23$\pm$0.05 &      \nodata    &      \nodata
\\ 0017-5129A      &   0.47$\pm$0.04 &   0.54$\pm$0.14 &
0.18$\pm$0.10 &      \nodata    &      \nodata    &      \nodata
\\ 0017-5129D      &   0.29$\pm$0.04 &      \nodata    &
0.30$\pm$0.07 &      \nodata    &      \nodata    &      \nodata
\\ 0018-4854A      &   0.68$\pm$0.01 &   0.77$\pm$0.01 &
0.91$\pm$0.01 &   0.11$\pm$0.01 &   0.62$\pm$0.01 &
0.80$\pm$0.01\\ 0018-4854B      &   1.02$\pm$0.02 & 1.35$\pm$0.03
&   1.33$\pm$0.02 &   0.41$\pm$0.02 & 0.92$\pm$0.03 &
0.40$\pm$0.03\\ 0018-4854C      & 0.14$\pm$0.03 &   2.04$\pm$0.03
&   0.43$\pm$0.03 & 0.04$\pm$0.05 &   0.11$\pm$0.03 &
0.81$\pm$0.03\\ 0018-4854D &   0.13$\pm$0.02 &   1.96$\pm$0.02 &
0.49$\pm$0.02 & \nodata    &      \nodata    &   0.81$\pm$0.02\\
0034-2250B &   0.38$\pm$0.02 &   1.99$\pm$0.02 &   0.62$\pm$0.02 &
0.04$\pm$0.04 &   0.09$\pm$0.02 &   0.89$\pm$0.02\\ 0034-2250C &
0.60$\pm$0.02 &   2.57$\pm$0.04 &   0.31$\pm$0.04 & \nodata    &
\nodata    &   0.52$\pm$0.03\\ 0035-3357A &   0.68$\pm$0.04 &
1.48$\pm$0.08 &   0.70$\pm$0.06 & \nodata    &      \nodata    &
\nodata   \\ 0035-3357C &   0.40$\pm$0.03 &   1.17$\pm$0.07 &
0.48$\pm$0.05 & 0.10$\pm$0.05 &   0.93$\pm$0.07 & 0.76$\pm$0.07\\
0035-3357D &   0.16$\pm$0.02 &   2.45$\pm$0.02 & 0.39$\pm$0.03 &
0.06$\pm$0.03 &   0.08$\pm$0.01 & 0.71$\pm$0.01\\ 0040-2350A &
3.05$\pm$0.01 &   7.14$\pm$0.02 & 1.29$\pm$0.02 & 0.73$\pm$0.02 &
0.28$\pm$0.01 & 0.30$\pm$0.01\\ 0040-2350B &   0.94$\pm$0.01 &
2.49$\pm$0.02 & 0.18$\pm$0.03 & 0.10$\pm$0.02 &   0.47$\pm$0.03 &
0.28$\pm$0.03\\ 0040-2350C &   0.63$\pm$0.02 &      \nodata    &
0.39$\pm$0.06 & 0.05$\pm$0.06 &      \nodata    &      \nodata \\
0102-4714B &   0.49$\pm$0.02 &      \nodata    & 0.48$\pm$0.04 &
0.04$\pm$0.06 &      \nodata    &      \nodata \\ 0105-1744A &
0.70$\pm$0.01 &   1.30$\pm$0.01 & 0.53$\pm$0.02 & 0.10$\pm$0.02 &
0.21$\pm$0.01 & 0.99$\pm$0.01\\ 0105-1744Ac &   0.30$\pm$0.00 &
1.68$\pm$0.01 & 0.26$\pm$0.01 & 0.03$\pm$0.01 &   0.08$\pm$0.01 &
0.47$\pm$0.01\\ 0105-1744B &   0.45$\pm$0.02 &   0.45$\pm$0.06 &
0.25$\pm$0.07 & 0.03$\pm$0.07 &   0.56$\pm$0.06 & 1.67$\pm$0.06\\
0105-1744D &   0.00$\pm$0.02 &   1.97$\pm$0.05 & 0.40$\pm$0.07 &
\nodata    &      \nodata    &      \nodata   \\ 0116-4439A &
0.70$\pm$0.01 &   1.09$\pm$0.02 &   0.73$\pm$0.01 & 0.13$\pm$0.02
&   0.53$\pm$0.02 &   0.49$\pm$0.02\\ 0116-4439B & 0.74$\pm$0.03 &
\nodata    &      \nodata    & \nodata    & \nodata    & \nodata
\\ 0116-4439D &   0.89$\pm$0.03 & \nodata    & 0.17$\pm$0.09 &
0.11$\pm$0.05 &      \nodata    & \nodata   \\ 0118-3624A &
0.67$\pm$0.01 &   1.38$\pm$0.06 & 0.22$\pm$0.04 & \nodata    &
\nodata    &   1.67$\pm$0.05\\ 0118-3624B & 1.01$\pm$0.03 &
8.50$\pm$0.08 &   0.83$\pm$0.05 & 0.20$\pm$0.05 &   0.28$\pm$0.08
&   0.09$\pm$0.08\\ 0122-3819C & 0.46$\pm$0.02 &   0.44$\pm$0.08 &
\nodata    & \nodata    & \nodata    & 2.99$\pm$0.08\\ 0146-2721A
&   0.84$\pm$0.02 & 0.12$\pm$0.07 & 0.43$\pm$0.03 & \nodata    &
\nodata    & 7.80$\pm$0.07\\ 0146-2721C & 0.65$\pm$0.02 &
0.64$\pm$0.04 & 0.83$\pm$0.03 & 0.09$\pm$0.04 & 0.49$\pm$0.04 &
1.38$\pm$0.04\\ 0146-2721E & 0.30$\pm$0.05 & 1.40$\pm$0.09 &
0.24$\pm$0.09 & \nodata    & \nodata    & \nodata \\ 0146-4855B &
0.68$\pm$0.02 & 1.42$\pm$0.07 & \nodata    & \nodata    & \nodata
& \nodata   \\ 0146-4855C & 0.30$\pm$0.03 & 2.22$\pm$0.04 &
0.36$\pm$0.08 & 0.09$\pm$0.04 & 0.27$\pm$0.03 & 1.01$\pm$0.03\\
0147-3506B & 0.36$\pm$0.02 & 0.17$\pm$0.07 & 0.13$\pm$0.05 &
\nodata    & \nodata    & 1.37$\pm$0.08\\ 0156-5629B &
0.60$\pm$0.02 & \nodata    & 0.44$\pm$0.05 & 0.03$\pm$0.08 &
\nodata    & \nodata   \\ 0156-5629D & 0.41$\pm$0.07 & \nodata &
0.55$\pm$0.10 & \nodata & \nodata    & \nodata   \\ 0227-4312E &
0.63$\pm$0.05 & 0.71$\pm$0.17 & 0.35$\pm$0.10 & 0.03$\pm$0.17 &
0.59$\pm$0.14 & \nodata \\ 0328-4807A & 0.50$\pm$0.02 & \nodata &
0.15$\pm$0.06 & \nodata & \nodata    &      \nodata \\ 0328-4807B
& 0.84$\pm$0.01 & 0.27$\pm$0.05 &   0.37$\pm$0.03 & 0.13$\pm$0.02
& 4.48$\pm$0.05 & 2.39$\pm$0.05\\ 0328-4807C & 0.17$\pm$0.05 &
\nodata & 0.40$\pm$0.07 & \nodata    & \nodata & \nodata \\
0414-5559C & 0.49$\pm$0.01 & \nodata    & 0.33$\pm$0.02 & \nodata
& \nodata    & \nodata   \\ 0414-5559D & 0.22$\pm$0.01 &
1.37$\pm$0.01 & 0.54$\pm$0.02 & 0.02$\pm$0.03 & 0.07$\pm$0.01 &
0.81$\pm$0.01\\ 0421-5624B & 0.65$\pm$0.01 & 4.75$\pm$0.02 &
0.37$\pm$0.03 & 0.16$\pm$0.02 & 0.15$\pm$0.02 & 0.22$\pm$0.02\\
0421-5624F & 0.29$\pm$0.04 & 1.33$\pm$0.13 & 0.35$\pm$0.07 &
0.04$\pm$0.10 &   0.66$\pm$0.08 & \nodata \\ 0426-4249C &
0.38$\pm$0.02 &   0.95$\pm$0.04 & 0.40$\pm$0.04 & 0.04$\pm$0.05 &
0.35$\pm$0.05 & 0.70$\pm$0.05\\ 0427-4802A & 2.34$\pm$0.02 &
8.43$\pm$0.07 & 1.25$\pm$0.03 & 0.46$\pm$0.02 & 1.32$\pm$0.05 &
0.24$\pm$0.05\\ 0427-4802B & 0.62$\pm$0.01 & 1.59$\pm$0.02 &
0.26$\pm$0.02 & 0.02$\pm$0.04 & 0.12$\pm$0.03 & 0.31$\pm$0.03\\
0435-5131B & 0.36$\pm$0.04 & 1.06$\pm$0.11 & 0.41$\pm$0.06 &
0.10$\pm$0.06 & 1.54$\pm$0.10 & 0.99$\pm$0.10\\ 0435-5131C &
1.37$\pm$0.04 & \nodata    & \nodata & \nodata    & \nodata    &
\nodata   \\ 0456-2141A & 0.70$\pm$0.03 & \nodata & \nodata &
\nodata    & \nodata & \nodata   \\ 0456-2141C & 0.57$\pm$0.04 &
2.23$\pm$0.07 & 0.18$\pm$0.11 & 0.08$\pm$0.09 & 0.20$\pm$0.08 &
0.24$\pm$0.08\\ 0533-5057A & 0.54$\pm$0.03 & \nodata    &
0.17$\pm$0.11 & \nodata & \nodata    & \nodata   \\ 0533-5057C &
0.21$\pm$0.03 & 1.64$\pm$0.03 & 0.47$\pm$0.04 & 0.03$\pm$0.08 &
0.05$\pm$0.03 & 0.62$\pm$0.03\\ 0533-5057D & 0.23$\pm$0.05 &
1.34$\pm$0.06 & 0.34$\pm$0.09 & \nodata    & \nodata    &
1.05$\pm$0.06\\ 0608-4734B & 1.16$\pm$0.02 & \nodata &
1.04$\pm$0.03 & 0.29$\pm$0.03 & 0.20$\pm$0.03 & 0.20$\pm$0.03\\
0608-4734D & 2.21$\pm$0.05 & 12.34$\pm$0.08 & 1.20$\pm$0.07 &
\nodata    & \nodata    & 0.16$\pm$0.05\\ 2113-4235B &
0.36$\pm$0.02 & \nodata & 0.32$\pm$0.05 & \nodata    & \nodata &
\nodata   \\ 2114-2302B & 0.62$\pm$0.03 & \nodata    & \nodata &
0.13$\pm$0.05 & \nodata    & \nodata   \\ 2114-2302C &
0.66$\pm$0.06 & \nodata & \nodata    & \nodata    & \nodata    &
\nodata   \\ 2114-2302D & 0.37$\pm$0.03 & \nodata    & \nodata &
\nodata    & \nodata & \nodata   \\ 2122-4245D & 0.34$\pm$0.08 &
\nodata    & \nodata & \nodata    & \nodata    & \nodata   \\
2123-2259A & 0.78$\pm$0.03 & \nodata    & 0.45$\pm$0.06 &
0.07$\pm$0.08 & \nodata    & \nodata \\ 2123-2259B & 0.41$\pm$0.02
& 0.38$\pm$0.06 & 0.26$\pm$0.04 & 0.03$\pm$0.05 & 0.88$\pm$0.06 &
\nodata \\ 2123-2259C & 0.32$\pm$0.03 & 0.58$\pm$0.07 &
0.40$\pm$0.05 & 0.09$\pm$0.05 &   1.20$\pm$0.07 & 1.86$\pm$0.07\\
2124-2314A & 0.93$\pm$0.02 &  10.57$\pm$0.06 & 0.89$\pm$0.03 &
0.04$\pm$0.07 & 0.06$\pm$0.05 &      \nodata \\ 2124-2314C &
0.41$\pm$0.05 & \nodata    &   0.48$\pm$0.09 & \nodata    &
\nodata    & \nodata \\ 2124-2314D & 0.36$\pm$0.02 & 0.58$\pm$0.06
& 0.43$\pm$0.04 & 0.08$\pm$0.04 & 1.14$\pm$0.05 & 2.09$\pm$0.05\\
2128-4614B & 0.39$\pm$0.02 & \nodata & 0.35$\pm$0.04 &
0.06$\pm$0.05 & \nodata    & \nodata \\ 2128-4614D & 0.43$\pm$0.02
& 0.62$\pm$0.08 & 0.41$\pm$0.04 & 0.03$\pm$0.06 & 1.01$\pm$0.07 &
\nodata \\ 2146-5935B & 0.52$\pm$0.03 & \nodata & 0.22$\pm$0.07 &
0.13$\pm$0.05 & \nodata    & \nodata \\ 2146-5935C & 0.47$\pm$0.03
& \nodata & 0.25$\pm$0.07 & 0.08$\pm$0.07 & \nodata & \nodata \\
2200-2048C & 0.34$\pm$0.02 & \nodata    & 0.41$\pm$0.04 & \nodata
& \nodata & \nodata   \\ 2211-4616A & 0.90$\pm$0.01 & \nodata &
0.89$\pm$0.02 & \nodata    & \nodata & 1.44$\pm$0.03\\ 2211-4616C
& 0.51$\pm$0.01 & 0.82$\pm$0.01 & 0.73$\pm$0.01 & 0.05$\pm$0.02 &
0.21$\pm$0.01 & 1.44$\pm$0.01\\ 2211-4616D & 0.43$\pm$0.01 &
0.66$\pm$0.04 & 0.57$\pm$0.02 & 0.03$\pm$0.04 & 0.50$\pm$0.03 &
1.48$\pm$0.03\\ 2211-4616E & 0.26$\pm$0.03 & 2.47$\pm$0.04 &
0.54$\pm$0.04 & \nodata    & \nodata    & \nodata   \\ 2211-4616F
& 0.17$\pm$0.02 & 1.94$\pm$0.02 & 0.36$\pm$0.02 & \nodata    &
\nodata    & 0.70$\pm$0.01\\ 2213-2743B &   0.31$\pm$0.02 &
0.43$\pm$0.05 & 0.40$\pm$0.04 & 0.05$\pm$0.05 &   0.59$\pm$0.05 &
2.48$\pm$0.05\\ 2226-6111B & 0.72$\pm$0.04 &   0.81$\pm$0.12 &
0.45$\pm$0.07 & \nodata    & \nodata    &   1.84$\pm$0.11\\
2226-6111C & 0.29$\pm$0.03 & \nodata    &   0.40$\pm$0.05 &
\nodata    & \nodata    & \nodata \\ 2315-4241A & 1.19$\pm$0.01 &
7.17$\pm$0.02 & 1.44$\pm$0.01 & 0.34$\pm$0.01 & 0.60$\pm$0.01 &
0.41$\pm$0.01\\ 2315-4241B & 0.93$\pm$0.00 & 1.67$\pm$0.01 &
0.96$\pm$0.01 & 0.15$\pm$0.01 & 0.99$\pm$0.01 & 1.07$\pm$0.01\\
2315-4241C & 0.59$\pm$0.01 & 0.13$\pm$0.03 & 0.62$\pm$0.01 &
\nodata    & \nodata    & 3.97$\pm$0.03\\ 2315-4241D &
0.71$\pm$0.01 & 2.21$\pm$0.02 & 0.87$\pm$0.01 & 0.06$\pm$0.02 &
0.19$\pm$0.01 &   0.99$\pm$0.01\\ 2316-2259A & 0.65$\pm$0.01 &
0.11$\pm$0.07 & 0.29$\pm$0.04 & 0.08$\pm$0.03 & 5.76$\pm$0.08 &
4.23$\pm$0.08\\ 2316-2259B & 0.33$\pm$0.02 & 1.09$\pm$0.04 &
0.41$\pm$0.03 & 0.07$\pm$0.03 & 0.60$\pm$0.04 & 0.53$\pm$0.04\\
2316-2259C & 0.15$\pm$0.05 & 2.17$\pm$0.06 & \nodata    & \nodata
& \nodata    & 1.37$\pm$0.04\\ 2353-6101A &   0.46$\pm$0.06 &
\nodata    & \nodata    & \nodata    & \nodata & \nodata \\
2353-6101B & 0.50$\pm$0.02 & \nodata    & 0.76$\pm$0.03 & \nodata
& \nodata & \nodata   \\
\enddata
\end{deluxetable}

\newpage
\begin{deluxetable}{lcccccccccc}
\footnotesize \tablecaption{Spectral characteristics of SCG's
emission-line galaxies. \label{tbl-3}} \tablewidth{0pt}
\tablehead{ \colhead{SCG \#} &\colhead{H$\alpha$/H$\beta$}
&\colhead{E(B-V)} & \colhead{$\frac{{\rm [NII]}}{{\rm H}\alpha}$}
& \colhead{$\frac{{\rm [OIII]}}{{\rm H}\beta}$} &
\colhead{$\frac{{\rm [SII]}}{{\rm H}\alpha}$} &
\colhead{$\frac{{\rm [OI]}}{{\rm H}\alpha}$}   &
\colhead{$\frac{{\rm [OI]}}{{\rm [OIII]}}$} &\colhead{$\frac{{\rm
[OII]}}{{\rm [OIII]}}$}   & \colhead{EW(H$\alpha$)}
&\colhead{Class.} \\
\colhead{}&\colhead{}&\colhead{}&\colhead{}&\colhead{}&\colhead{}&\colhead{}&\colhead{}&\colhead{}&\colhead{(\AA)}&
\colhead{} } \startdata 0004-5044D      &  6.79 & 0.70 &  -0.36
&\nodata &  -0.40 &  -0.60 &\nodata &\nodata &    7.3 &  H:  \\
0017-5129A      & 17.11 & 1.44 &  -0.34 &  -0.35 &  -0.79 &\nodata
&\nodata &\nodata &   18.9 &  H   \\ 0017-5129D      &  0.00 &
0.00 &  -0.54 &\nodata &  -0.53 &\nodata &\nodata &\nodata & 12.3
&  H   \\ 0018-4854A      &  4.27 & 0.26 &  -0.17 &  -0.13 & -0.05
&  -0.94 &  -0.32 &   0.04 &   25.2 &  L \\ 0018-4854B & 3.03 &
0.00 &   0.01 &   0.13 &   0.12 &  -0.39 &  -0.04 & -0.40 & 7.7 &
L \\ 0018-4854C      &  5.69 & 0.55 &  -0.87 & 0.28 & -0.39 &
-1.38 &  -1.20 &   0.20 &   48.0 &  H   \\ 0018-4854D & 4.84 &
0.36 &  -0.89 &   0.27 &  -0.33 &\nodata &\nodata & 0.10 & 48.2 &
L:\\ 0034-2250B      &  4.27 & 0.32 &  -0.42 & 0.28 & -0.22 &
-1.37 &  -1.20 &   0.12 &   11.3 & H:  \\ 0034-2250C & 10.11 &
1.02 &  -0.23 &   0.35 &  -0.55 &\nodata &\nodata & 0.25 & 10.1 &
H   \\ 0035-3357A      & 6.84 & 0.64 &  -0.17 & 0.13 & -0.18
&\nodata &\nodata &\nodata & 3.1 &  L \\ 0035-3357C & 10.40 & 0.97
&  -0.40 &   0.01 & -0.36 &  -0.92 &  -0.44 & 0.39 &   18.1 & L:\\
0035-3357D & 3.18 & 0.08 &  -0.78 &   0.38 &  -0.42 & -1.21 &
-1.14 & -0.11 & 64.1 &  H:  \\ 0040-2350A &  2.78 & 0.00 & 0.48 &
0.85 & 0.11 &  -0.14 &  -0.55 &  -0.52 &    2.9 & Sy2\\ 0040-2350B
& 12.04 & 1.09 &  -0.03 &   0.33 & -0.79 & -0.94 & -0.78 & 0.02 &
13.5 &  L:\\ 0040-2350C      & 0.00 & 0.00 & -0.20 &\nodata &
-0.41 &  -1.31 &\nodata &\nodata & 6.0 & H?  \\ 0102-4714B      &
0.00 & 0.00 &  -0.31 &\nodata & -0.32 &  -1.37 &\nodata &\nodata &
19.4 &  H?  \\ 0105-1744A & 2.60 & 0.00 & -0.16 &   0.11 &  -0.27
&  -0.98 &  -0.68 & 0.00 & 25.1 &  L \\ 0105-1744Ac     &  4.21 &
0.31 & -0.52 & 0.21 & -0.60 &  -1.50 & -1.25 &  -0.16 &  290.3 & H
\\ 0105-1744B & 7.72 & 0.80 &  -0.35 &  -0.40 &  -0.64 &  -1.44 &
-0.58 & 0.64 & 16.5 &  H   \\ 0105-1744D      &  4.32 & 0.33
&\nodata & 0.27 & -0.41 &\nodata &\nodata &\nodata &   22.1 &  H
\\ 0116-4439A &  4.55 & 0.31 & -0.16 &   0.02 &  -0.15 &  -0.88 &
-0.40 & -0.15 &   28.0 &  L \\ 0116-4439B      &  0.00 & 0.00 &
-0.13 &\nodata &\nodata &\nodata &\nodata &\nodata &    4.1 & L?\\
0116-4439D      & 0.00 & 0.00 &  -0.05 &\nodata &  -0.78 & -0.96
&\nodata &\nodata & 6.6 &  L \\ 0118-3624A      & 26.87 & 1.80 &
-0.18 &   0.03 & -0.73 &\nodata &\nodata &   1.17 & 7.4 & H: \\
0118-3624B      & 12.00 & 1.09 &   0.00 &   0.86 & -0.12 & -0.63 &
-1.00 &  -0.48 & 8.6 & Sy2\\ 0122-3819C & 20.74 & 1.60 & -0.35 &
-0.45 &\nodata &\nodata &\nodata & 1.32 &   11.2 &  H? \\
0146-2721A      & 4.55 & 0.37 &  -0.08 & -0.95 & -0.38 &\nodata
&\nodata &   1.09 & 3.4 &  H:  \\ 0146-2721C &  3.61 & 0.12 &
-0.19 &  -0.20 & -0.08 &  -1.05 &  -0.36 & 0.21 & 9.7 & L \\
0146-2721E      & 9.11 & 0.93 &  -0.52 & 0.09 & -0.65 &\nodata
&\nodata &\nodata & 16.6 & H \\ 0146-4855B & 21.97 & 1.58 & -0.18
&   0.06 &\nodata &\nodata &\nodata &\nodata & 5.4 &  L?\\
0146-4855C & 6.74 & 0.69 &  -0.52 &   0.30 & -0.47 &  -1.01 &
-0.86 &   0.37 &   18.7 & H:  \\ 0147-3506B &  9.26 & 0.95 & -0.45
& -0.84 & -0.93 &\nodata &\nodata & 0.64 &   19.1 &  H   \\
0156-5629B & 0.00 & 0.00 & -0.22 &\nodata &  -0.36 &  -1.51
&\nodata &\nodata & 20.2 &  L:\\ 0156-5629D      &  0.74 & 0.00 &
-0.39 &\nodata & -0.26 &\nodata &\nodata &\nodata &    4.4 & L:\\
0227-4312E & 13.61 & 1.26 & -0.20 &  -0.23 &  -0.50 & -1.43 &
-0.75 &\nodata &   20.8 &  H \\ 0328-4807A      & 0.00 & 0.00 &
-0.30 &\nodata &  -0.82 &\nodata &\nodata &\nodata &   11.8 &  H
\\ 0328-4807B &  9.16 & 0.87 & -0.08 & -0.62 &  -0.46 & -0.82 &
0.29 &   0.84 &    8.4 &  L:\\ 0328-4807C      &  0.00 & 0.00 &
-0.77 &\nodata &  -0.40 &\nodata &\nodata &\nodata & 20.6 &  H \\
0414-5559C &  0.00 & 0.00 & -0.31 &\nodata & -0.48 &\nodata
&\nodata &\nodata & 4.4 & H   \\ 0414-5559D &  3.82 & 0.23 & -0.66
& 0.12 & -0.28 &  -1.59 &  -1.26 & 0.03 &   23.0 &  H:  \\
0421-5624B &  4.65 & 0.33 & -0.19 & 0.66 &  -0.44 &  -0.78 & -0.95
& -0.48 &   20.0 & Sy1$_N$\\ 0421-5624F      & 24.01 & 1.71 &
-0.55 &   0.02 & -0.52 &  -1.33 & -0.89 &\nodata &   11.1 &  H \\
0426-4249C &  7.57 & 0.78 & -0.43 &  -0.07 &  -0.43 & -1.31 &
-0.78 & 0.26 &   38.6 &  H \\ 0427-4802A      & 24.03 & 1.65 &
0.36 & 0.83 &   0.04 & -0.23 & -0.56 &   0.25 &    7.9 &  Sy2\\
0427-4802B      &  9.84 & 1.00 & -0.21 & 0.14 &  -0.62 &  -1.63 &
-1.32 &   0.01 & 21.9 &  H   \\ 0435-5131B & 17.03 & 1.44 & -0.45
&  -0.06 & -0.44 &  -0.92 & -0.41 & 0.75 &   24.6 &  H: \\
0435-5131C &  0.00 & 0.00 &   0.14 &\nodata &\nodata &\nodata
&\nodata &\nodata & 1.3 &  L?\\ 0456-2141A      &  0.00 & 0.00 &
-0.16 &\nodata &\nodata &\nodata &\nodata &\nodata &    3.3 & L?\\
0456-2141C &  5.94 & 0.59 & -0.25 &   0.31 &  -0.76 &  -1.08 &
-0.94 & -0.31 &   16.1 &  H: \\ 0533-5057A      &  0.00 & 0.00 &
-0.27 &\nodata &  -0.76 &\nodata &\nodata &\nodata &    5.1 &  H
\\ 0533-5057C      & 3.40 & 0.14 &  -0.69 &   0.21 &  -0.34 &
-1.58 &  -1.33 &  -0.13 & 32.3 &  H:  \\ 0533-5057D      & 4.18 &
0.31 &  -0.64 &   0.11 & -0.48 &\nodata &\nodata &   0.18 & 41.0 &
H   \\ 0608-4734B & 0.00 & 0.00 &   0.06 &\nodata & 0.02 & -0.54 &
-0.71 &  -0.70 & 4.5 &  L \\ 0608-4734D & 4.72 & 0.34 & 0.34 &
1.07 & 0.07 &\nodata &\nodata & -0.61 & 1.2 &  Sy2\\ 2113-4235B &
0.00 & 0.00 &  -0.44 &\nodata & -0.49 &\nodata &\nodata &\nodata &
7.6 & H   \\ 2114-2302B &  0.00 & 0.00 & -0.21 &\nodata &\nodata &
-0.88 &\nodata &\nodata &    3.0 & L:\\ 2114-2302C & 0.00 & 0.00 &
-0.18 &\nodata &\nodata &\nodata &\nodata &\nodata & 9.0 & L?\\
2114-2302D      & 0.00 & 0.00 &  -0.43 &\nodata &\nodata &\nodata
&\nodata &\nodata & 7.3 &  H?  \\ 2122-4245D & 0.00 & 0.00 & -0.47
&\nodata &\nodata &\nodata &\nodata &\nodata & 3.1 & H   \\
2123-2259A &  7.43 & 0.70 &  -0.11 &\nodata & -0.37 & -1.09
&\nodata &\nodata &    5.1 &  L \\ 2123-2259B & 10.86 & 1.07 &
-0.40 & -0.49 &  -0.63 &  -1.45 & -0.50 &\nodata & 15.6 & H \\
2123-2259C      &  7.49 & 0.78 & -0.50 &  -0.29 & -0.43 & -0.99 &
-0.24 &   0.68 & 14.5 & H: \\ 2124-2314A      & 15.66 & 1.30 &
-0.04 &   0.95 & -0.10 & -1.34 & -1.79 &\nodata & 7.1 & Sy2\\
2124-2314C &  0.00 & 0.00 & -0.38 &\nodata &  -0.32 &\nodata
&\nodata &\nodata & 6.4 & H:  \\ 2124-2314D      &  8.78 & 0.84 &
-0.45 &  -0.29 & -0.40 & -1.07 & -0.29 &   0.76 &   36.5 & L:\\
2128-4614B &  0.00 & 0.00 &  -0.41 &\nodata &  -0.45 & -1.25
&\nodata &\nodata &   14.5 &  H \\ 2128-4614D      & 18.46 & 1.50
& -0.37 &  -0.30 &  -0.44 & -1.37 & -0.62 &\nodata & 17.0 & H \\
2146-5935B      & 10.64 & 1.06 & -0.29 &\nodata & -0.70 & -0.81
&\nodata &\nodata & 14.1 &  H   \\ 2146-5935C & 0.00 & 0.00 &
-0.32 &\nodata & -0.60 &  -1.12 &\nodata &\nodata &   12.6 &  H
\\ 2200-2048C & 14.15 & 1.29 & -0.48 &\nodata & -0.43 &\nodata
&\nodata &\nodata & 10.1 &  H \\ 2211-4616A &  0.00 & 0.00 & -0.04
&\nodata & -0.05 &\nodata &\nodata & 0.16 &    1.2 & L \\
2211-4616C &  3.35 & 0.13 & -0.29 &  -0.09 &  -0.14 &  -1.27 &
-0.72 & 0.23 &   24.2 &  H: \\ 2211-4616D &  9.55 & 0.97 & -0.37 &
-0.24 &  -0.28 & -1.41 & -0.71 &   0.68 &   12.3 &  H: \\
2211-4616E      & 6.93 & 0.65 & -0.59 &   0.35 &  -0.29 &\nodata
&\nodata &\nodata &   17.5 & L:\\ 2211-4616F & 4.26 & 0.32 & -0.78
&   0.27 &  -0.45 &\nodata &\nodata & 0.02 &   45.2 & H \\
2213-2743B &  5.45 & 0.52 & -0.51 & -0.40 & -0.41 & -1.30 & -0.44
& 0.67 &   18.9 &  H   \\ 2226-6111B & 11.96 & 1.15 & -0.15 &
-0.16 &  -0.39 &\nodata &\nodata & 0.87 &    5.3 & H   \\
2226-6111C      &  0.00 & 0.00 &  -0.53 &\nodata &  -0.40 &\nodata
&\nodata &\nodata &   31.5 & H \\ 2315-4241A      & 12.40 & 1.12 &
0.07 &   0.79 &   0.12 & -0.39 &  -0.69 &   0.20 & 4.1 & Sy2\\
2315-4241B      & 11.38 & 1.05 & -0.04 &   0.16 & -0.06 & -0.77 &
-0.44 &   0.58 & 6.3 & L \\ 2315-4241C      & 4.60 & 0.32 & -0.23
&  -0.91 & -0.22 &\nodata &\nodata &   0.77 & 10.0 & L:\\
2315-4241D &  7.66 & 0.73 & -0.15 &   0.30 &  -0.09 &  -1.21 &
-1.02 & 0.38 &    4.1 & L:\\ 2316-2259A      &  7.85 & 0.75 &
-0.19 & -1.01 &  -0.57 & -1.05 &   0.45 &   1.02 & 11.1 &  L:\\
2316-2259B      &  9.52 & 0.97 & -0.49 &  -0.02 & -0.42 & -1.10 &
-0.62 &   0.24 &   37.3 & H \\ 2316-2259C &  7.86 & 0.81 &  -0.82
&   0.29 &\nodata &\nodata &\nodata & 0.57 & 18.3 & H?  \\
2353-6101A      &  0.00 & 0.00 & -0.34 &\nodata &\nodata &\nodata
&\nodata &\nodata & 4.1 &  H? \\ 2353-6101B &  0.00 & 0.00 & -0.30
&\nodata &  -0.12 &\nodata &\nodata &\nodata & 6.0 & L:\\
\enddata
\end{deluxetable}

\clearpage
\begin{deluxetable}{rcccc}
\footnotesize \tablecaption{Criteria for spectral
classification\label{tbl-4}} \tablewidth{0pt} \tablehead{
\colhead{Class} & \colhead{$log(\frac{{\rm [OIII]}}{{\rm
H}\alpha})$}  & \colhead{$log(\frac{{\rm [OI]}}{{\rm H}\alpha})$}
& \colhead{$log(\frac{{\rm [SII]}}{{\rm H}\alpha})$} &
\colhead{$log(\frac{{\rm [NII]}}{{\rm H}\alpha})$} } \startdata
Starburst  & any & $< -1.1$   & $< -4.0$   & $< -2.2$    \\
Seyfert    & $>0.4$ & $\ge -1.1$ & $\ge -4.0$ & $\ge -2.2$ \\
LINER      & $<0.4$ & $\ge -1.1$ & $\ge -4.0$ & $\ge -2.2$ \\
\enddata
\end{deluxetable}

\newpage
\begin{deluxetable}{lcccccccc}
\footnotesize \tablecaption{Activity types per group in
SCG\label{tbl-5}} \tablewidth{0pt} \tablehead{ \colhead{Group
id.}&\colhead{in/out} &\colhead{dAGN} &\colhead{Sy1}
&\colhead{Sy2}     & \colhead{LINER}      &\colhead{SFG}
&\colhead{EMLG} &\colhead{No em.} } \startdata 0004-5044 & 4/4 &
B,C    &\nodata& \nodata  &\nodata  & D     &\nodata  &  A
\\ 0017-5129 & 3/5 &  C      &\nodata& \nodata  &\nodata  & A,D
&\nodata  & \nodata   \\ 0018-4854 & 4/4 & \nodata &\nodata&
\nodata  & A,B,D   & C     & \nodata & \nodata   \\ 0034-2250 &
3/4 &  A      &\nodata& \nodata  &\nodata  & B,C   & \nodata &
\nodata   \\ 0035-3357 & 4/4 & \nodata &\nodata& \nodata  & A,C &
D     & \nodata &  B        \\ 0040-2350 & 4/5 &  A &\nodata&  Ac
& B       & C     &\nodata  & \nodata   \\ 0054-5312 & 6/6 & C,D,E
&\nodata& \nodata  &\nodata &\nodata&\nodata  &  A,B,F \\
0100-2208 & 3/4 &  C &\nodata& \nodata  &\nodata &\nodata&\nodata
&  A,D      \\ 0102-4714 & 4/4 &  C,D &\nodata& \nodata  &\nodata
& B &\nodata  &  A        \\ 0105-1744 & 4/5 & \nodata &\nodata&
\nodata  & A       & Ac,B,D&\nodata  & \nodata   \\ 0106-4722 &
4/4 & A,B,C,D&\nodata& \nodata  &\nodata  &\nodata&\nodata  &
\nodata \\ 0111-3203 & 4/4 &  A,B,C,D&\nodata& \nodata &\nodata
&\nodata&\nodata  & \nodata   \\ 0116-4439 & 3/4 & \nodata
&\nodata& \nodata  & A,B,D   &\nodata&\nodata  & \nodata \\
0118-3624 & 3/5 & \nodata &\nodata&    B     & \nodata & A &     C
& \nodata   \\ 0122-3819 & 4/4 &  B      &\nodata& \nodata
&\nodata  & C     &\nodata  &  A,D      \\ 0146-2721 & 4/5 &  D
&\nodata& \nodata  & C       & A,E   &\nodata  & \nodata   \\
0146-4855 & 3/4 &  A      &\nodata& \nodata  & B & C     &\nodata
& \nodata   \\ 0147-3506 & 3/5 &  A &\nodata& \nodata  &\nodata &
B     & C       & \nodata   \\ 0156-5629 & 4/4 &  A &\nodata&
\nodata  & B,D &\nodata&\nodata  & C         \\ 0227-4312 & 6/6 &
D &\nodata& \nodata  &\nodata  & E     & B &  A,C,F    \\
0242-1750 & 3/5 &  A,B    &\nodata& \nodata &\nodata  &\nodata& C
& \nodata   \\ 0328-4807 & 3/4 & \nodata &\nodata& \nodata  &  B &
A,C   &\nodata  & \nodata   \\ 0414-5559 & 4/4 &  A &\nodata&
\nodata  &\nodata  & C,D   &\nodata &  B        \\ 0421-5624 & 4/7
& \nodata & B     & \nodata &\nodata  & F &\nodata  & A,Ac \\
0426-4249 & 4/5 &  A &\nodata& \nodata  &\nodata  & C &\nodata  &
B,Ac     \\ 0427-4802 & 3/6 & \nodata &\nodata&  A &\nodata  & B
&\nodata  &  C \\ 0435-5131 & 4/4 &  D &\nodata& \nodata  & C & B
&\nodata  &  A        \\ 0456-2141 & 4/6 &  B,D &\nodata& \nodata
& A       & C &\nodata  & \nodata   \\ 0533-5057 & 3/4 & \nodata
&\nodata& \nodata  &\nodata  & A,C,D &\nodata  & \nodata \\
0608-4734 & 6/6 & \nodata &\nodata&    D & B &\nodata&\nodata &
A,C,E,F  \\ 2113-4235 & 4/4 &  A &\nodata& \nodata  &\nodata & B
&\nodata &  C,D      \\ 2114-2302 & 4/6 & \nodata &\nodata&
\nodata  & B,C &   D &\nodata  & A         \\ 2114-4840 & 3/5 & A
&\nodata& \nodata  &\nodata  &\nodata&\nodata  &  B,E \\ 2122-4245
& 5/6 &  A      &\nodata& \nodata  &\nodata  & D &\nodata  & B,C,F
\\ 2123-2259 & 3/4 & \nodata &\nodata& \nodata  & A & B,C
&\nodata  & \nodata   \\ 2123-6105 & 4/5 & A &\nodata& \nodata
&\nodata  &\nodata& E       &  B,C      \\ 2124-2314 & 4/4 &
\nodata &\nodata&  A       & D       & C &\nodata  &  B \\
2128-4614 & 5/5 &  A      &\nodata& \nodata &\nodata  & B,D & C,E
& \nodata   \\ 2146-5935 & 3/4 &  A &\nodata& \nodata  &\nodata &
B,C   &\nodata  & \nodata   \\ 2200-2048 & 4/4 &  A,D &\nodata&
\nodata &\nodata  & C &\nodata &  B \\ 2211-4616 & 6/6 &  B
&\nodata& \nodata  & A,E & C,D,F &\nodata  & \nodata   \\
2213-2743 & 4/4 &  C &\nodata& \nodata &\nodata  & B &\nodata  &
A,D      \\ 2226-3546 & 5/5 & \nodata &\nodata& \nodata  &\nodata
&\nodata& C,E     &  A,B,D \\ 2226-6111 & 5/6 & \nodata &\nodata&
\nodata &\nodata  & B,C &\nodata  & A,E,F    \\ 2315-4241 & 4/4 &
\nodata &\nodata&  A & B,C,D &\nodata&\nodata  & \nodata   \\
2316-2259 & 3/4 & \nodata &\nodata& \nodata  & A       & B,C
&\nodata  & \nodata \\ 2351-4015 & 5/5 & A       &\nodata& \nodata
&\nodata &\nodata&\nodata  &  Ac,B,C,D \\ 2353-6101 & 3/4 &
\nodata &\nodata& \nodata  & B       & A     &\nodata  &  C \\
2358-4339 & 4/5 & B       &\nodata& \nodata  &\nodata
&\nodata&\nodata  &  A,C,D     \\
\enddata
\end{deluxetable}

\clearpage
\begin{deluxetable}{rcccc}
\footnotesize \tablecaption{Frequency of activity types in
CGs\label{tbl-6}} \tablewidth{0pt} \tablehead{ \colhead{}    &
\colhead{HCG}  & \colhead{HCG}  & \colhead{SCG}  & \colhead{SCG}
\\ \colhead{}    & \colhead{group}& \colhead{core} &
\colhead{all}  & \colhead{$\geq$4m.} } \startdata No of Galaxies &
62    & 43    & 193  & 145   \\
              No em.      & 40\%  & 37\%  & 27\% & 32\% \\
              LLAGN       & 21\%  & 23\%  & 22\% & 24\% \\
               AGN        & 18\%  & 21\%  & 19\% & 18\% \\
               SFG        & 21\%  & 19\%  & 32\% & 26\% \\
\enddata
\end{deluxetable}

\clearpage
\begin{deluxetable}{rccc}
\footnotesize \tablecaption{Kolmogorov--Smirnov test on SCG galaxy
magnitudes\label{tbl-7}} \tablewidth{0pt} \tablehead{ \colhead{} &
\colhead{AGN}  & \colhead{NOEM}  & \colhead{SFG}} \startdata
              LLAGN       & 71.3\%  & 58.3\%  &  3.0\% \\
               AGN        & \nodata & 73.8\%  & 35.2\% \\
               NOEM       & \nodata & \nodata & 12.2\% \\
\enddata
\end{deluxetable}

\clearpage

\figcaption[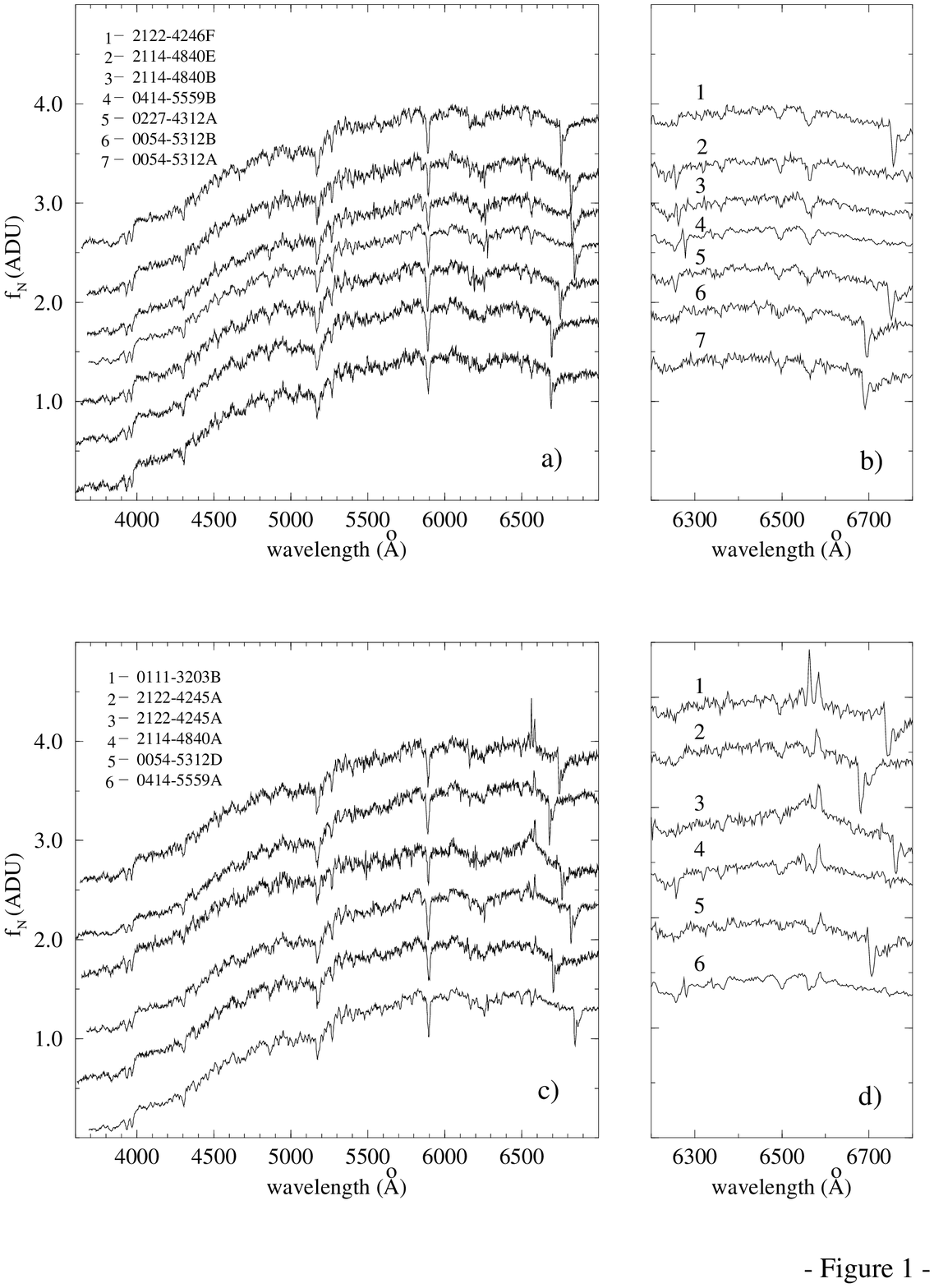]{Examples of different spectral types in
the SCG galaxies sample.  The spectra are presented in the rest
frame of the galaxies.  The fluxes are in ADU and normalized by
the mean counts.  They have been vertically shifted for
comparison.  In a) we show examples of galaxies without emission
lines.  In c) we show examples of low luminosity 
AGNs.  The figures on the right is a zoom on the red part of the
spectrum.  They show details of nitrogen, sulphur and H$\alpha$
emission.}

\figcaption[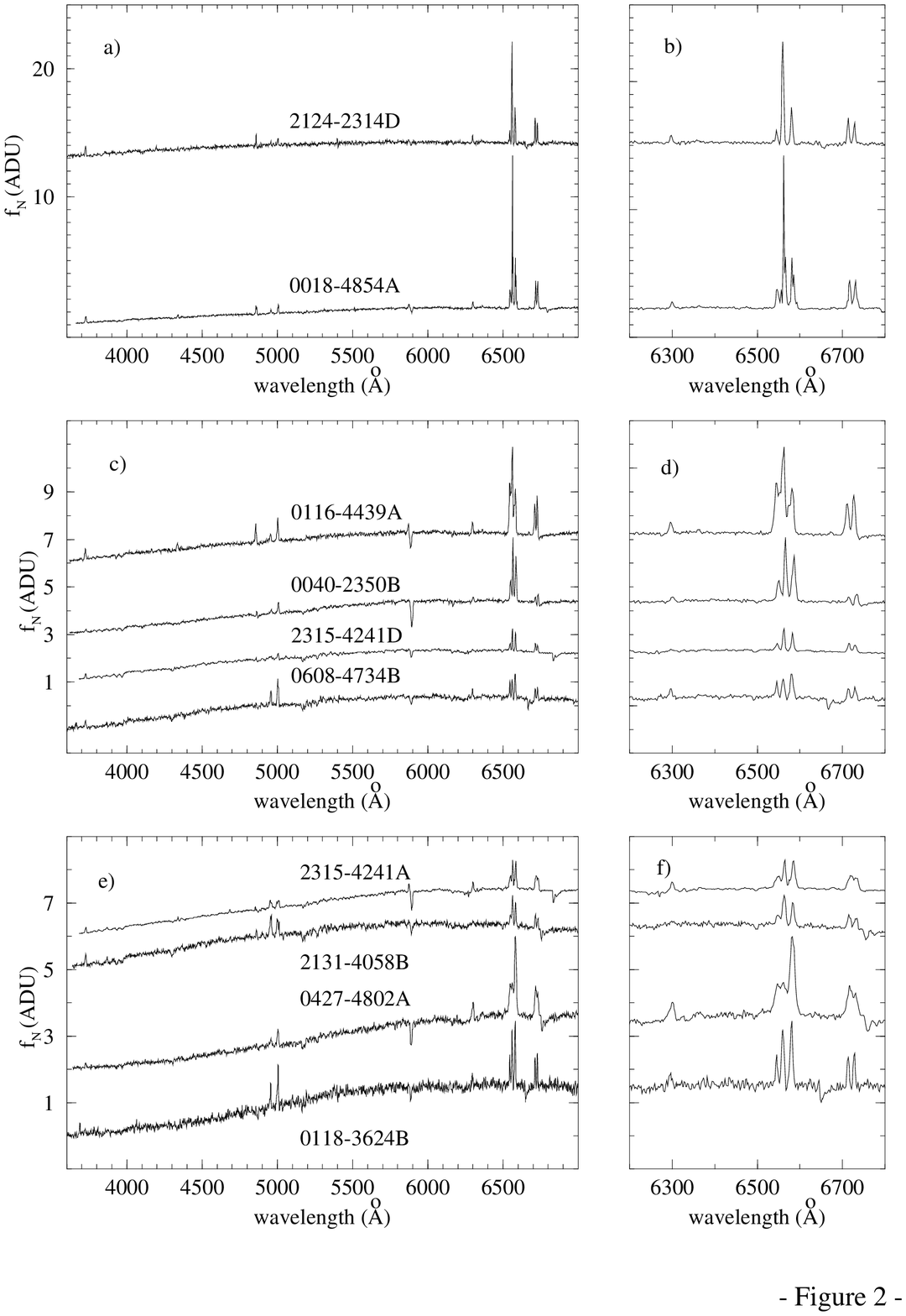]{Examples of different spectral types in
the SCG galaxies sample.  The wavelength and fluxes are as in
Figure~1.  In a) we show examples of galaxies with strong star
formation in their nucleus.  In c) we show examples of
star--forming galaxies with already decreasing star formation.
The figures to the right show how the nitrogen and sulphur
emission lines varied with star formation.}

\figcaption[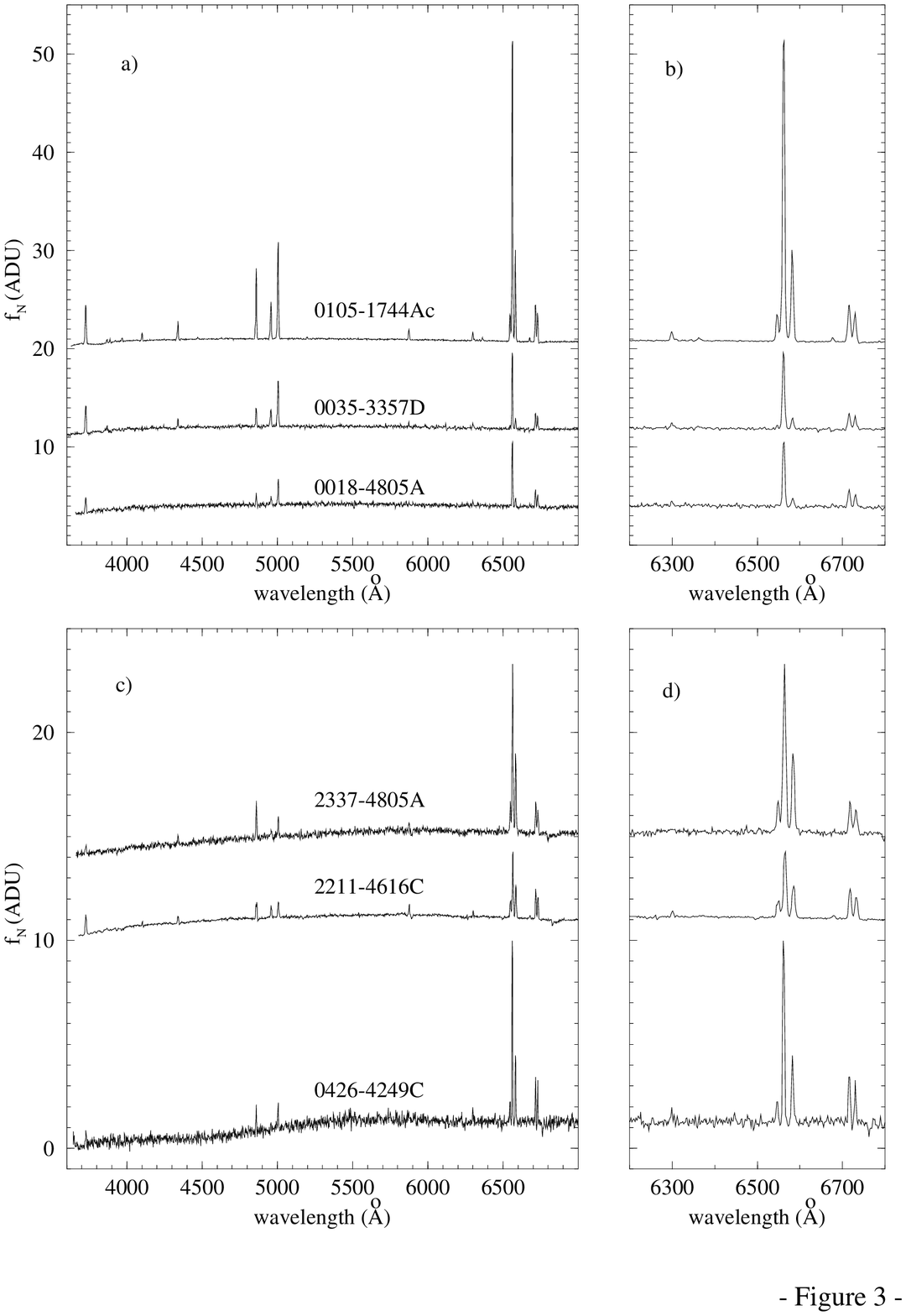]{Examples of different spectral types in
the SCG galaxies sample.  The wavelength and fluxes are as in
Figure~1 and 2.  In a) and c) we show different examples of
LINERs.  In e) we show examples of Seyfert 2 galaxies.  Notice how
the nitrogen and sulphur emission lines varied in these galaxies
as compared to star--forming galaxies.}

\figcaption[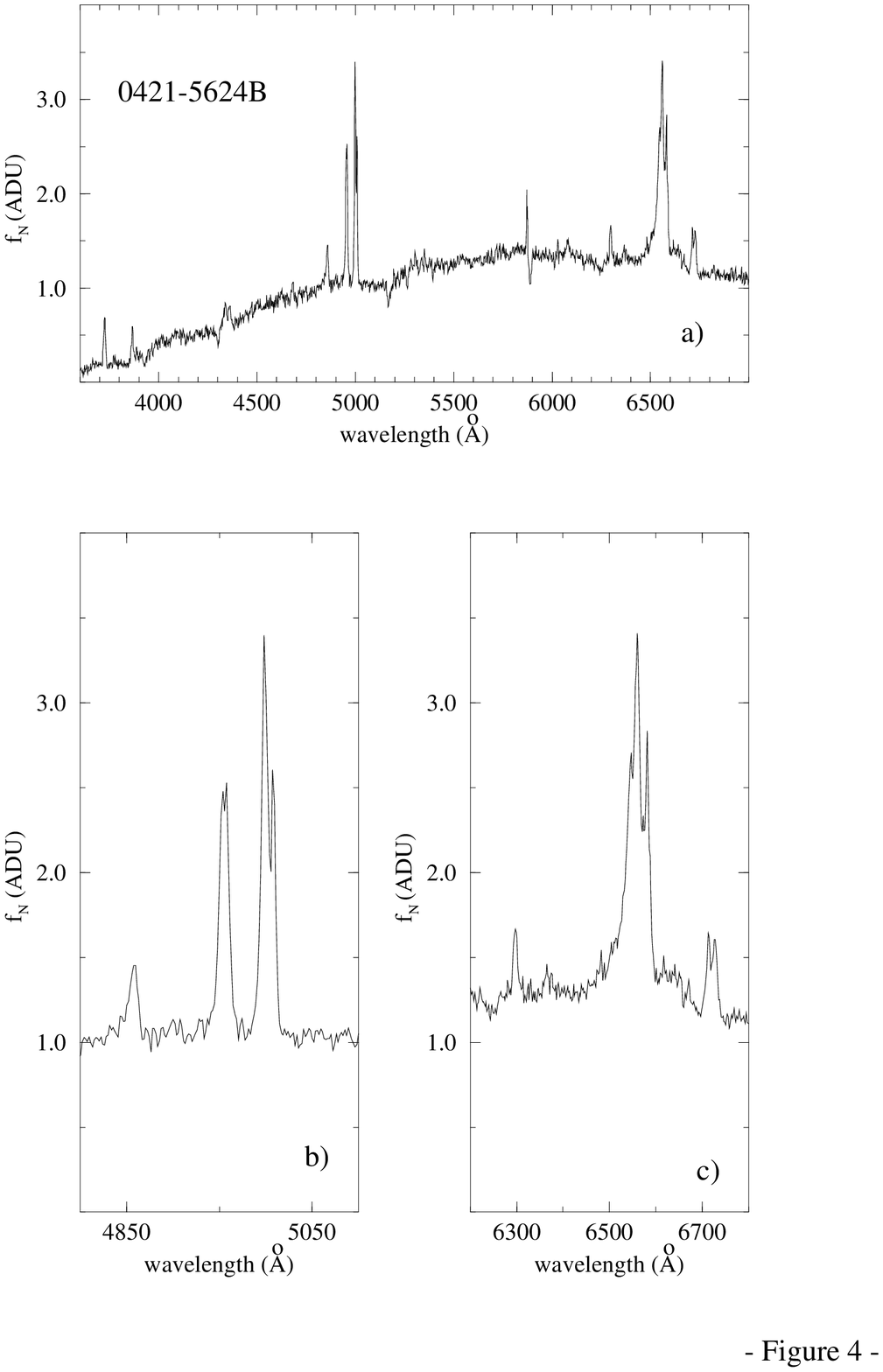]{The only example of a possible Seyfert
1 in our sample.  The wavelength and fluxes are as in the other
figures.  The Seyfert 1 nature is suggested by the possible large
base component in H$\alpha$ in c).  Although it is less visible,
this wide line component may also be present in H$\beta$.  The
double peak on the oxygen lines are visible in 3 different poses.
We could have more than one emission line regions in this galaxy.}

\figcaption[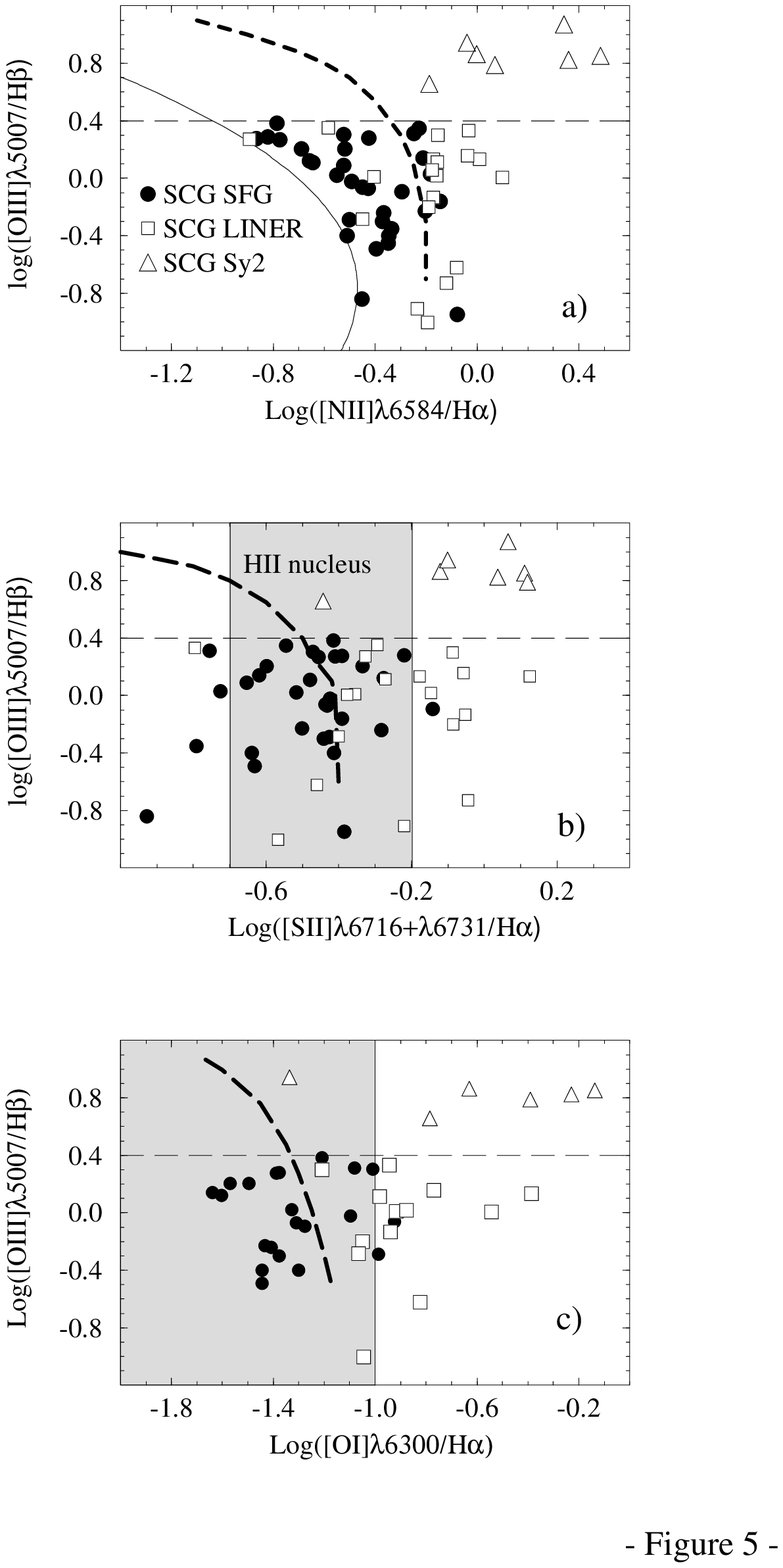]{The three emission line diagnostic
diagrams used to classify the activity types of the SCG galaxies.
In these figures the horizontal line at log([{\rm O
III}]/H$\alpha$)=0.4 separates high from low excitation galaxies
(Seyfert 2 from LINERs or {\rm H II} galaxies from SBGs).  The
dashed bold lines are empirical limits separating AGNs from
star--forming galaxies.  In a), the thin continuous line show the
locus occupied by normal {\rm H II} regions.  In b) and c) the
gray regions indicate typical values of line ratios encountered in
{\rm H II} Nucleus Galaxies.}

\figcaption[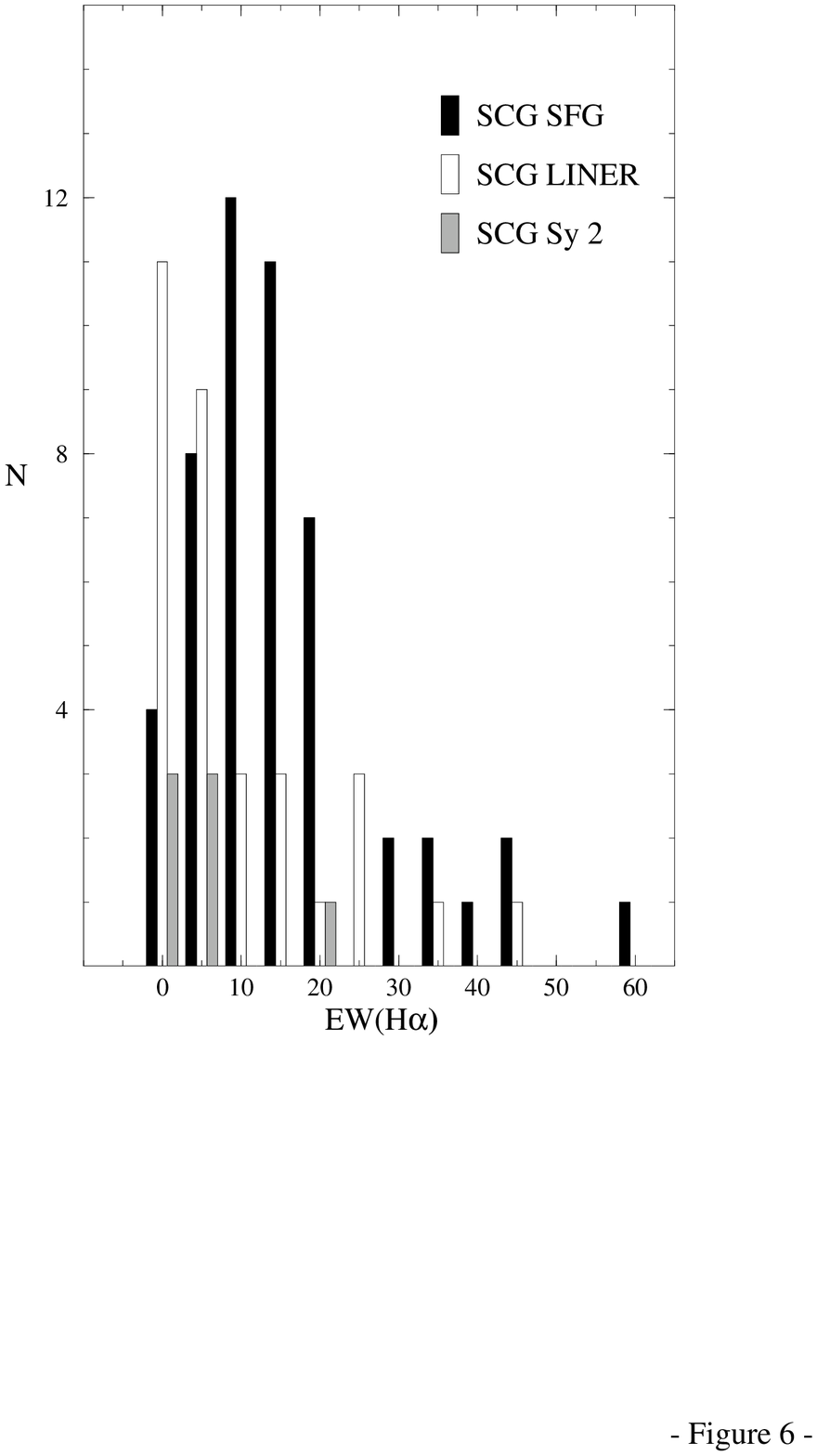]{Distribution of H$\alpha$ equivalent
width for the emission--line galaxies in the SCG sample. The
relatively low values suggest low star formation rates, confirming
the classification of these galaxies as {\rm H II} Nucleus
Galaxies.}

\figcaption[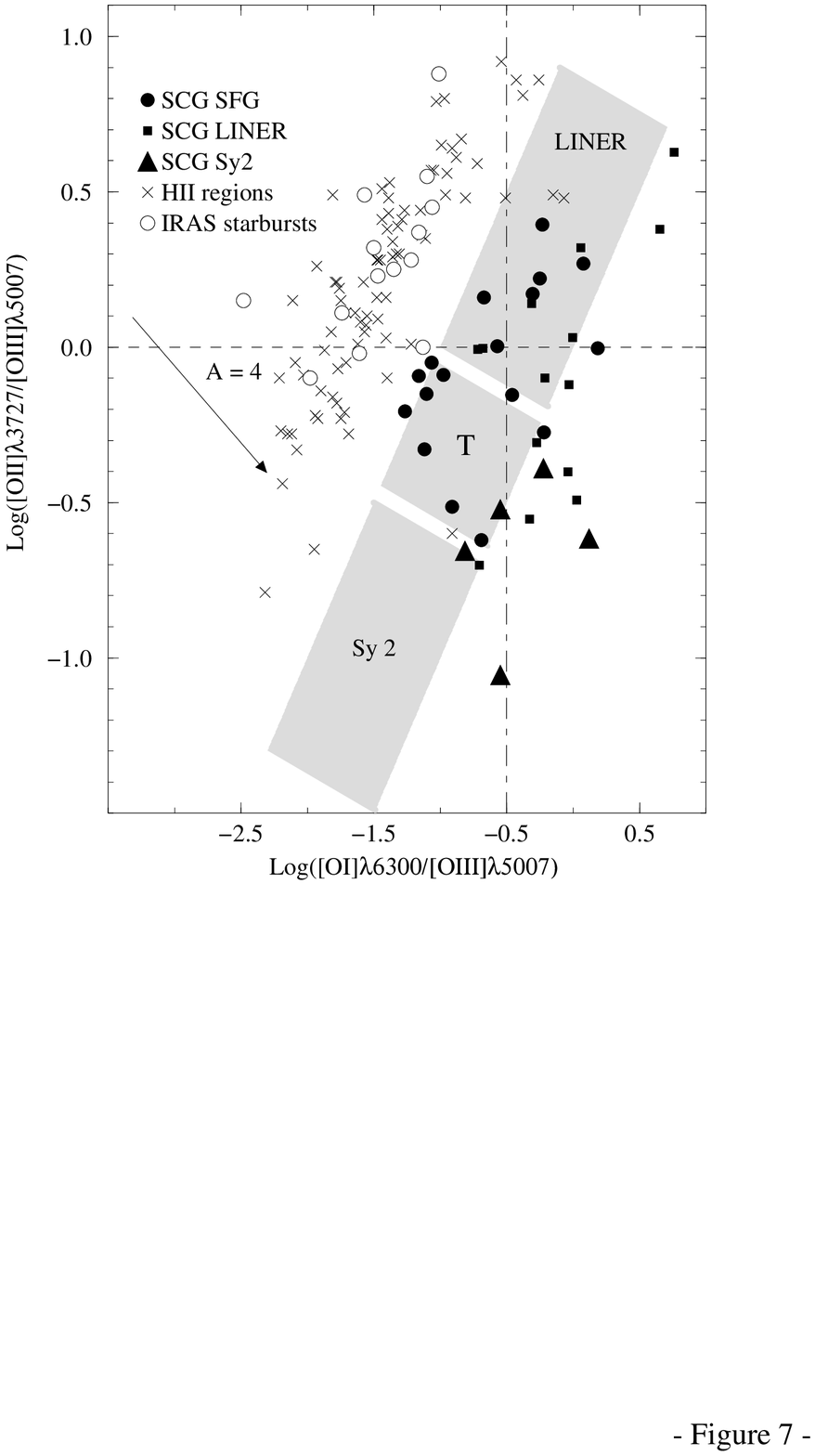]{Diagnostic diagram used by Heckman
(1980) to define the LINERs.  The gray regions are the loci
occupied by Seyfert 2 and LINERs in Heckman's sample.  The
vertical and horizontal lines are the limits proposed by Heckman
to separate these two types.  The line ratios in this figure are
not corrected for extinction.  The vector corresponds to an
extinction A$_V = 4$ magnitudes, which is much higher than what is
observed in our galaxies.}

\figcaption[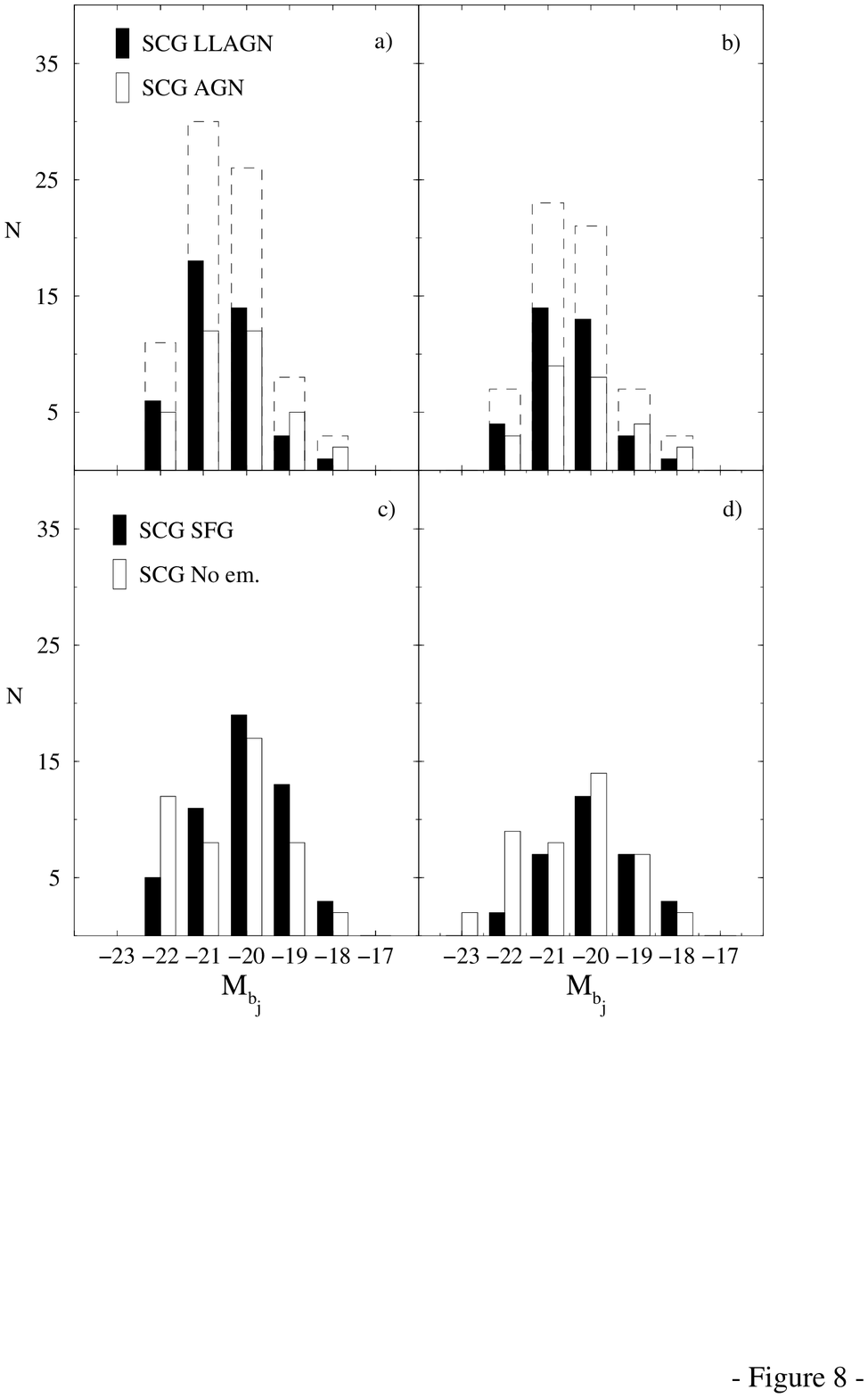]{Distribution of the luminosities for
the galaxies with different activity types.  The left panels show
the result for the whole sample while the right panels include
only SCGs with more than 3 galaxies.}

\figcaption[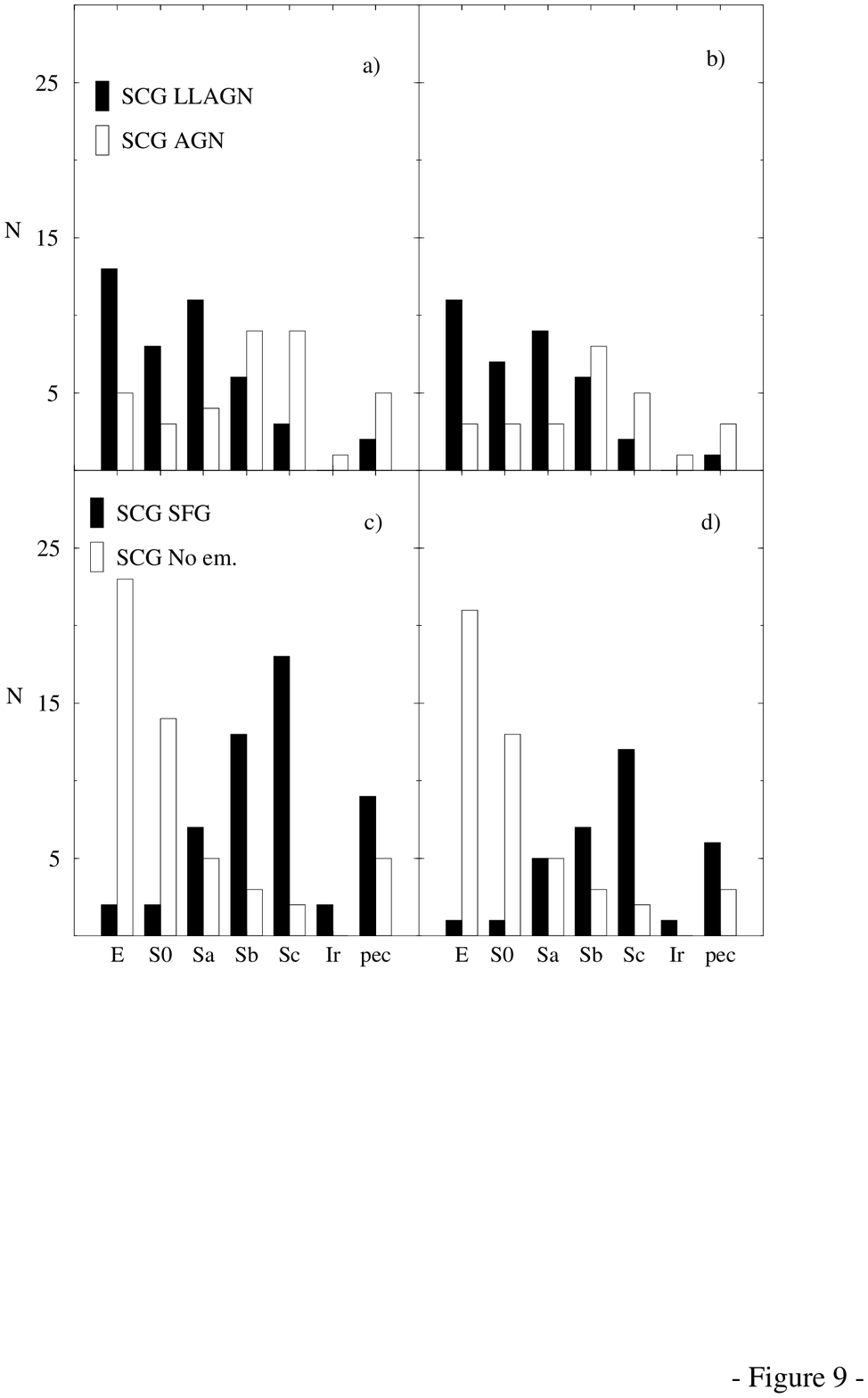]{Distribution of the morphologies of the
galaxies with different activity types.  Like in Figure~8, the
left panels include the whole sample while the right panels
include only SCGs with more than 3 galaxies.  The morphological p
type corresponds to peculiar morphologies, obviously related to
cases of merging galaxies.}

\end{document}